\UseRawInputEncoding

\documentclass[runningheads]{llncs}
\usepackage{geometry}
\usepackage{graphicx}
\usepackage{graphicx}
\usepackage{enumitem}
\usepackage{amsmath}
\usepackage{breqn}
\usepackage{pbox}
\usepackage{multirow}
\usepackage{url}
\usepackage{xurl}

\usepackage{hyperref}
\usepackage{flushend}
\usepackage{color}
\usepackage{fixltx2e}
\usepackage{balance}
\usepackage{caption}
\usepackage{stfloats}
\usepackage{booktabs}
\usepackage[table,xcdraw]{xcolor}
\graphicspath{{figures/}}
\usepackage[framemethod=TikZ]{mdframed}
\usepackage{subfigure}
\usepackage{setspace}

\usepackage{listings}
\definecolor{bluebell}{rgb}{0.64, 0.64, 0.82}
\usepackage{listings,xcolor}
\begin{document}
\title{FedDICE: A ransomware spread detection in a distributed integrated clinical environment using federated learning and SDN based mitigation}
%
%
\author{Chandra Thapa\inst{1} \and
Kallol Krishna Karmakar\inst{2} \and
Alberto Huertas Celdran\inst{3} \and
Seyit Camtepe\inst{1} \and
Vijay Varadharajan\inst{2} \and
Surya Nepal\inst{1,4}
}
\authorrunning{Thapa et al.}
%
\institute{CSIRO Data61, Australia \and
The University of Newcastle, Australia \and
University of Zurich, Switzerland \and
Cyber Security Cooperative Research Centre, Australia\\
\email{\{chandra.thapa, seyit.camtepe, surya.nepal\}@data61.csiro.au}\\
\email{\{kallolkrishna.karmakar, vijay.varadharajan\}@newcastle.edu.au}, \email{huertas@ifi.uzh.ch}
}


\titlerunning{FedDICE}

\maketitle              
%

\begin{abstract}
An integrated clinical environment (ICE) enables the connection and coordination of the internet of medical things around the care of patients in hospitals. However, ransomware attacks and their spread on hospital infrastructures, including ICE, are rising. Often the adversaries are targeting multiple hospitals with the same ransomware attacks. These attacks are detected by using machine learning algorithms. But the challenge is devising the anti-ransomware learning mechanisms and services under the following conditions: (1) provide immunity to other hospitals if one of them got the attack, (2) hospitals are usually distributed over geographical locations, and (3) direct data sharing is avoided due to privacy concerns. In this regard, this paper presents a federated distributed integrated clinical environment, aka. \emph{FedDICE}. FedDICE integrates federated learning (FL), which is privacy-preserving learning, to SDN-oriented security architecture to enable collaborative learning, detection, and mitigation of ransomware attacks. We demonstrate the importance of FedDICE in a collaborative environment with up to four hospitals and four popular ransomware families, namely WannaCry, Petya, BadRabbit, and PowerGhost. Our results find that in both IID and non-IID data setups, FedDICE achieves the centralized baseline performance that needs direct data sharing for detection. However, as a trade-off to data privacy, FedDICE observes overhead in the anti-ransomware model training, e.g., $28\times$ for the logistic regression model. Besides, FedDICE utilizes SDN's dynamic network programmability feature to remove the infected devices in ICE.

\end{abstract}
%
%
%

\section{Introduction}
\label{sec:introduction}

The latest advancement of computing paradigms and communications is influencing the revolution of many heterogeneous scenarios. Healthcare is one of the most relevant due to its impact on human well-being. Nowadays, hospitals are adapting their operating theaters and infrastructure with new paradigms, for example, the Internet of Medical Things (IoMT)~\cite{miot}, and Medical Cyber-Physical Systems (MCPS)~\cite{cyberphysical}, to enable open coordination and interoperability of heterogeneous medical devices (e.g., data logging device) and applications (e.g., clinical decision making). These infrastructures are called \textit{Integrated Clinical Environments (ICE)}~\cite{ICE_architecture} and hold the promise of providing innovative and optimized ways to monitor, diagnose, and treat patients. Moreover, ICE enables a holistic view of the ongoing condition of a patient and improves the patient's data collection system by creating the IoMT and MCPS around the care of patients. Refer to Fig.~\ref{fig:icefw} for the ICE functional architecture. 

However, hospitals, including their ICEs, are under cyberattacks. We have been witnessing how hospitals are impacted by multiple types of cyberattacks that expose sensitive data or disrupt critical tasks such as surgeries or treatments~\cite{attack1}. This problem is greatly influenced by having medical devices or computers without enough --or obsolete-- cybersecurity mechanisms and connection to the internet. In this context, from current hospitals to the next generation equipped with ICE, the expected healthcare evolution will aggravate the situation because ICE devices have not been designed to satisfy security requirements, and consequently, are vulnerable to cyberattacks such as ransomware. 

Ransomware cyberattacks deserve special attention from the research community as they are skyrocketing during this COVID-19 pandemic period in healthcare~\cite{ransomware_skyrocketing,ransomware_skyrocketing3}. For example, it is increased by 45\% since November 2020~\cite{ransomware_skyrocketing2}. However, they are not new events. They accounted for 85\% of all malware, and more than 70\% of attacks confirmed data disclosure based on Verizon in 2018~\cite{verizon_report}. In January 2018, the Hancock Health Hospital (US) paid attackers \$55,000 to unlock systems following a ransomware infection~\cite{ransomware_pay}. Considering all, it is evident that without addressing ransomware attacks impacting clinical scenarios, we cannot achieve the benefits of ICE in healthcare. 

For ransomware detection, various mechanisms have been developed. But, the traditional cybersecurity mechanisms based on signatures, e.g., SIDS~\cite{ids_survey}, and rule-based static policies, e.g., traditional policy-based detection in software-defined networks (SDN), are no longer suitable for detecting new ransomware families that have not been seen before or use encrypted communications. To solve this drawback, a vast number of solutions based on machine learning (ML) and deep learning (DL) have been proposed~\cite{ML_survey}. However, ML/DL requires sufficient ransomware data for training and testing its algorithms. Moreover, data should have a good quality~\cite{data_quality}, for example, all targeted classes (e.g., ransomware families) to enabling better ML/DL learning on their features. 

Unfortunately, it is not easy to find all required ransomware data in one hospital's ICE environment, as one hospital might not suffer from all ransomware families. Thus there is a need for collaboration among hospitals. The straightforward way is to collect the ransomware data directly from the multiple hospital's ICE and perform machine learning. This approach is called \emph{centralized learning} (CL). The challenge here is data privacy since the data from ICE is privacy critical as it can contain patients' health data and can expose the hospitals' internal ICE networks to others (refer to section~\ref{sec:data-privacy} for details). Also, privacy in ICE is recommended by NIST (NIST RISK MANAGEMENT FRAMEWORK, supra note 57, at v) frameworks~\cite{nist}. Thus hospitals are reluctant to collaborate with direct data sharing. Besides, the hospitals (ICEs) are separated and located at different geo-locations making the raw data sharing difficult. 

Overall, a collaborative healthcare framework is needed, where several hospitals (ICEs) work together to create a powerful anti-ransomware model with data privacy. As the collaborative framework has multiple ICEs (hospitals) located in different places, it is called distributed ICE (DICE). Each participating hospital with its local ICE environment can share its locally trained anti-ransomware models instead of raw ransomware data in DICE. They can use those shared models in two ways; (i) use individually, and (ii) aggregate models to form one global model. The former is not preferred because it is a burden to the hospitals as they need to keep, manage, and update these multiple models received from multiple hospitals with time.

In this regard, federated learning (FL) is a suitable candidate. FL aggregates the multiple models to form one global model and updates it with time by considering data privacy in a distributed machine learning. Thus this paper integrates FL to DICE, and the integrated framework is called \emph{FedDICE}. In ransomware attack scenarios, its spread is one of the major concerns in the connected environment~\cite{threat} such as ICE. So, the FedDICE is investigated by focusing on the detection and mitigation of ransomware spread. Specifically, this work presents the performance of FL in ransomware spread detection considering four popular families, namely WannaCry, Petya, BadRabbit and PowerGhost. This is not known yet despite FL's widespread use in other healthcare scenarios, such as diagnosing medical conditions from medical imaging like MRI scans and X-rays~\cite{reviewpaper_our,naturefed}. 
For the mitigation of the attacks in ICE, SDN policies are applied to separate the ransomware-infected system/device from the network.

For convenience, we list our overall contributions based on the research question (\textbf{RQ}) in the following: 
\begin{itemize} [leftmargin = 30px]
    \item [\textbf{RQ1}] \textbf{Need of a collaborative framework in ransomware spread detection:}
    We study how the models trained on one ransomware family, we call \emph{singly-trained models}, perform over other ransomware families. Our results demonstrate that the singly-trained models are not effective in general in detecting different ransomware. Thus collaboration is required if all ransomware data is not available at one hospital. Refer to section~\ref{sec:results_and_analysis} for details.  
    
    \item [\textbf{RQ2}] \textbf{Security architecture of FedDICE:} We present a security architecture of FedDICE for the model development, detection, and mitigation of ransomware spread. The architecture is based on the SDN functionalities and components. Refer to section~\ref{sec:proposed_solution} for details.
    
    \item [\textbf{RQ3}] \textbf{Performance of FedDICE over centralized learning:}
    This is an important research question highlighting the importance of the FL paradigm in DICE, i.e., FedDICE, both from the requirements and performance sides in ransomware spread detection. Our studies demonstrate that FL achieves the baseline performance, i.e., centralized learning, under both the IID and the non-IID ransomware data distribution in FedDICE. Moreover, FL is even better than centralized learning in some cases. Thus one can avoid using privacy-unfriendly centralized learning. Besides, our results show that a simple model such as logistic regression performs well in ransomware spread detection in our setup. Refer to section~\ref{sec:results_and_analysis} for details. 
 
    \item [\textbf{RQ4}] \textbf{Mitigation of the ransomware spread:} 
    As the mitigation of ransomware threat is equally important as its detection, we implement policy-based mitigation in the SDN-managed FedDICE. Refer to section~\ref{sec:mitigation} for details.
\end{itemize}


\section{Background}
\label{sec:background}
This section provides background on the various techniques and frameworks that we are using in this paper. This includes the ICE framework, DICE framework, centralized and federated learning.

\subsection{ICE framework}
\label{sec:ICE}

ICE consists of multiple medical and non-medical equipment (including IoMT), their connections, and control systems designed to enable a patient-centric infrastructure. To standardize the ICE framework for interoperability, the American Society for Testing and Materials (ASTM) states a high-level architecture for ICE~\cite{ICE_architecture}. The proposed framework is depicted in Fig.~\ref{fig:icefw}. The framework enables the internal and external medical or relevant devices to connect to the system via \emph{ICE equipment interface} and \emph{external interface}, respectively. All the devices/equipment are connected in a network, and it is managed and monitored by \emph{ICE network controller}. Thus, it has access to the network flow information of the devices in the ICE network. Such information is stored in \emph{data logger} for forensic analysis. For SDN-based infrastructure, the \emph{SDN controller} is kept at the same layer of the ICE network controller, and they operate collaboratively. The medical applications run on the \emph{ICE supervisor} platform. These applications use the data generated by medical devices for medical decisions, in which the devices are operated on the patient, and the clinician makes the decision.   


\subsection{DICE framework}
\label{sec:DICE}
\begin{figure}[t]
	\centering
		\subfigure[]{
			\includegraphics[trim=0.2cm 0.2cm 0.2cm 0.2cm, clip=true, width=0.35\linewidth]{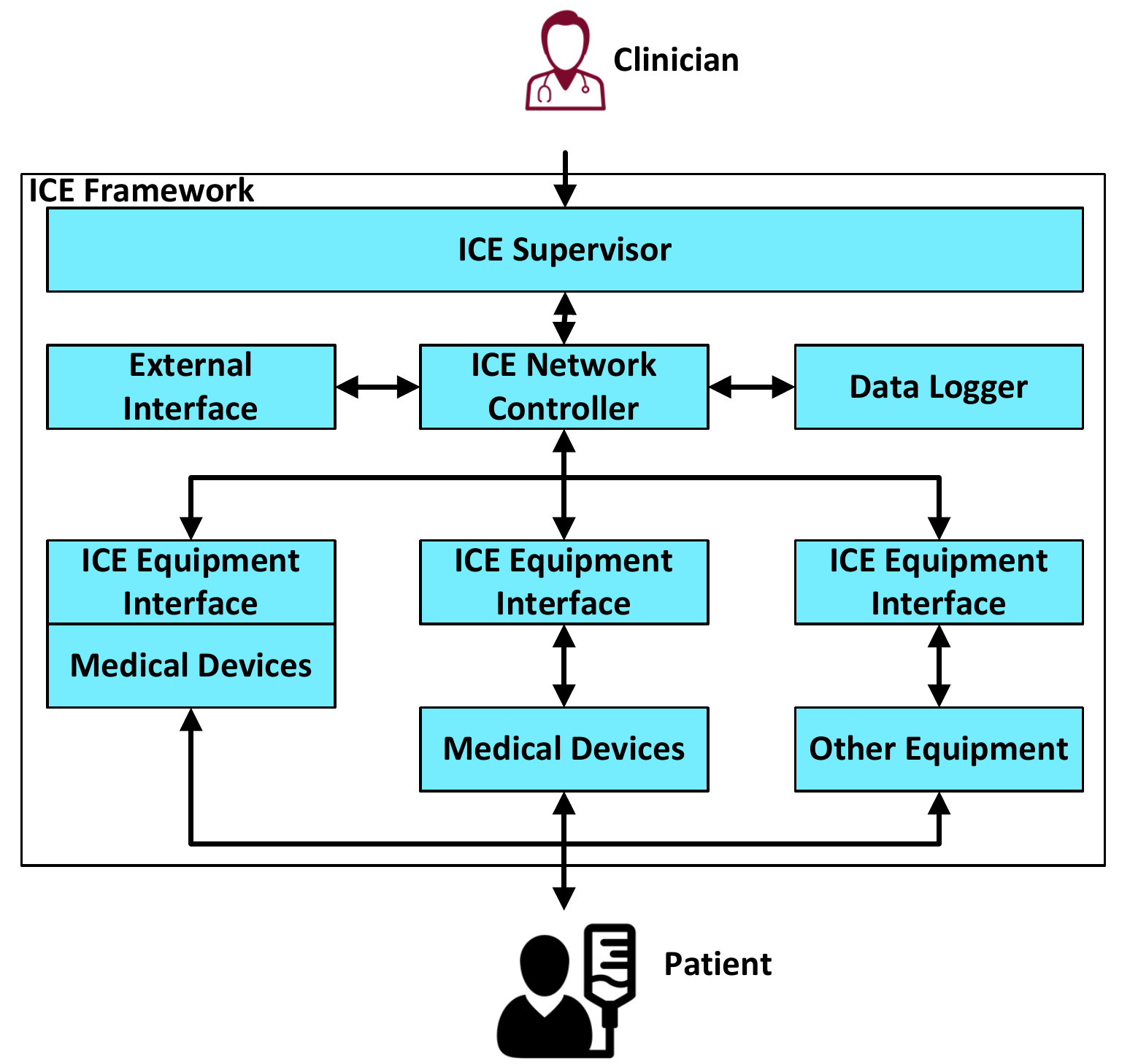}
			\label{fig:icefw}
		}
		\hskip-2pt
	\subfigure[]{
			\includegraphics[width=0.4\linewidth]{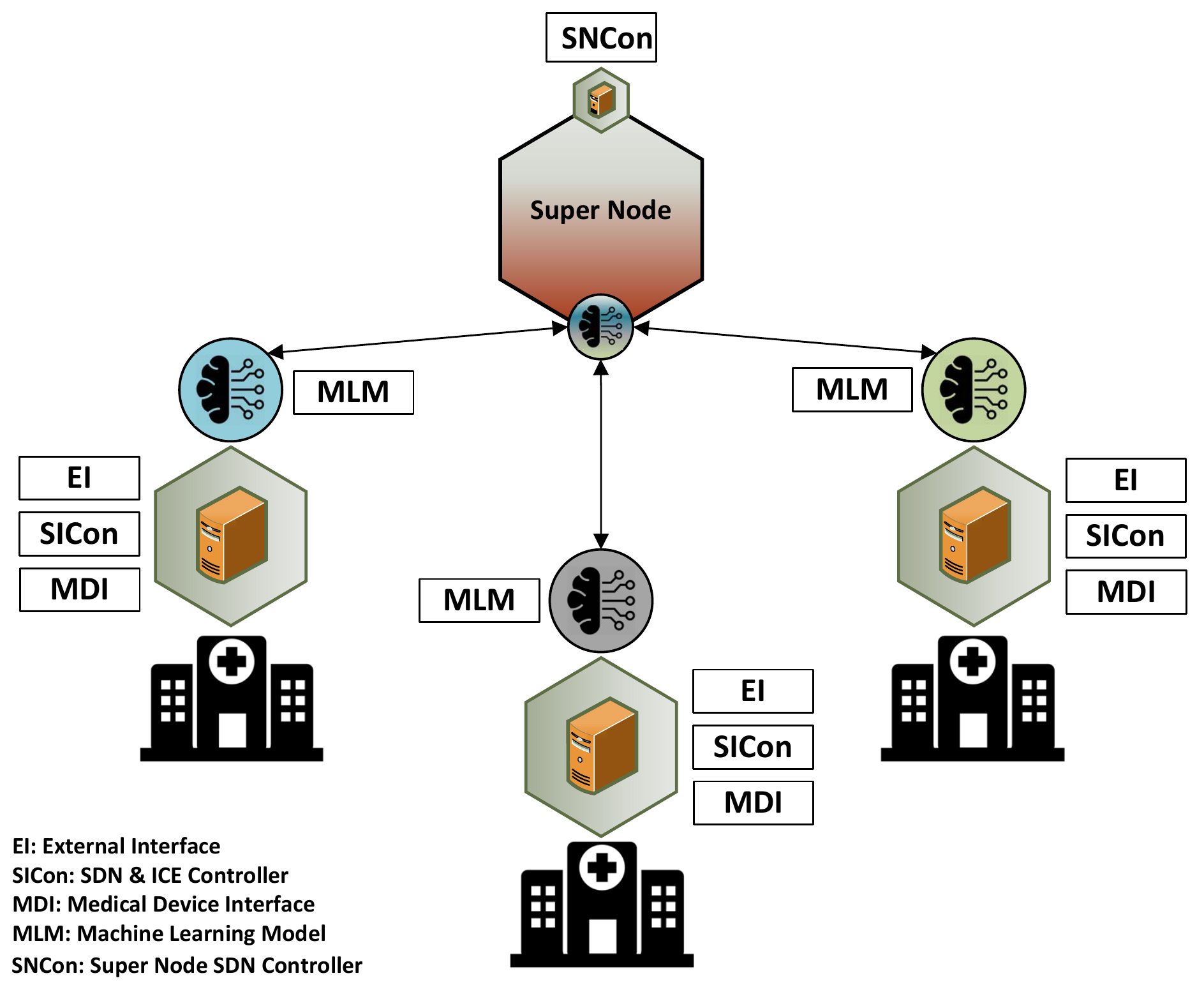}
			\label{fig:fed_learning} 
		}
		\caption{\small{(a) An ICE framework by ASTM Standard F2671-2009~\cite{ICE_architecture}, and (b) an example scenario of DICE with three hospitals. In DICE, each hospital has its local ICE with components such as external interface (EI), SDN and ICE controller (SICon), and medical device interface (MDI). The locally trained machine learning models (MLM) can be shared with the supernode, which also has an SDN controller (SNCon).}}
		\label{fig:}
\end{figure}

DICE is formed by connecting multiple ICE via a trusted party, called supernode. The supernode is connected through the external interface of each ICE. It has some computing power and is responsible only for enabling collaboration among the participating ICEs. A supernode can have an SDN controller to help with the routing of packets between hospitals. Supernode acts only following the commands from the participating ICEs. By default, neither it interferes with the decisions made by the local SDN controller (at each hospital) for its local ICE, nor it sees/collect or transfers data of the connected ICE without their approvals. Fig.~\ref{fig:fed_learning} illustrates an example scenario of DICE where three different hospitals at different regions are connected via a supernode to collaborating by sharing their local inferences for the improvement of their local ICE network security. Overall, DICE is better than a single standalone ICE for various aspects, including improved security via knowledge sharing (e.g., models), data sharing, and resource management.  


\textbf{DICE specific requirements:} This work is related to the security and privacy of DICE; thus, we further explore and list its requirements in the following: 
\begin{enumerate} 
    \item \textbf{Distributed computing:} The underlying structure of the DICE framework has distributed resources, including computation. Thus, the computations, including ML/AI anti-ransomware model training/testing and data pre-processing, are required to do in the local resources of ICEs. Moreover, the distributed computing enables reliability by localizing the possibility of the system failure, e.g., limiting fault to the local ICE, add scalability by allowing more ICEs to the framework, and faster computations due to distributed computing capability.   
    
    \item \textbf{Local control:} The local ICEs belong to independent hospitals, so the control over their local network and SDN policies execution must remain within. The outsider (e.g., supernode) can only facilitate them to route and share model, data, and knowledge for overall benefit but only with their consent.   
    
    \item \textbf{Collaborative approach:} DICE needs a collaborative approach to solve various issues, including security. For example, the participating ICEs can share their knowledge on detecting some ransomware to alert or improve the detection capability of other ICEs. Also, this approach enables better Ml/AI threat detection model training by providing access to more data types.  

    \item \textbf{Privacy:} Privacy is another important requirement demanded by regulations (e.g., General Data Protection Regulation of Europe~\cite{gdpr} and NIST~\cite{nist}) and users. In DICE, the information shared between the ICEs of different organizations needs to be protected; the approach should be privacy-by-design and privacy-by-default.  
\end{enumerate}
\subsection{Centralized learning and federated learning}
\label{sec:centralized_learning}

\begin{figure}[t!]
	\centering
		\subfigure[]{
			\includegraphics[trim=11.2cm 3cm 2.5cm 0.5cm, clip=true, width=0.33\linewidth]{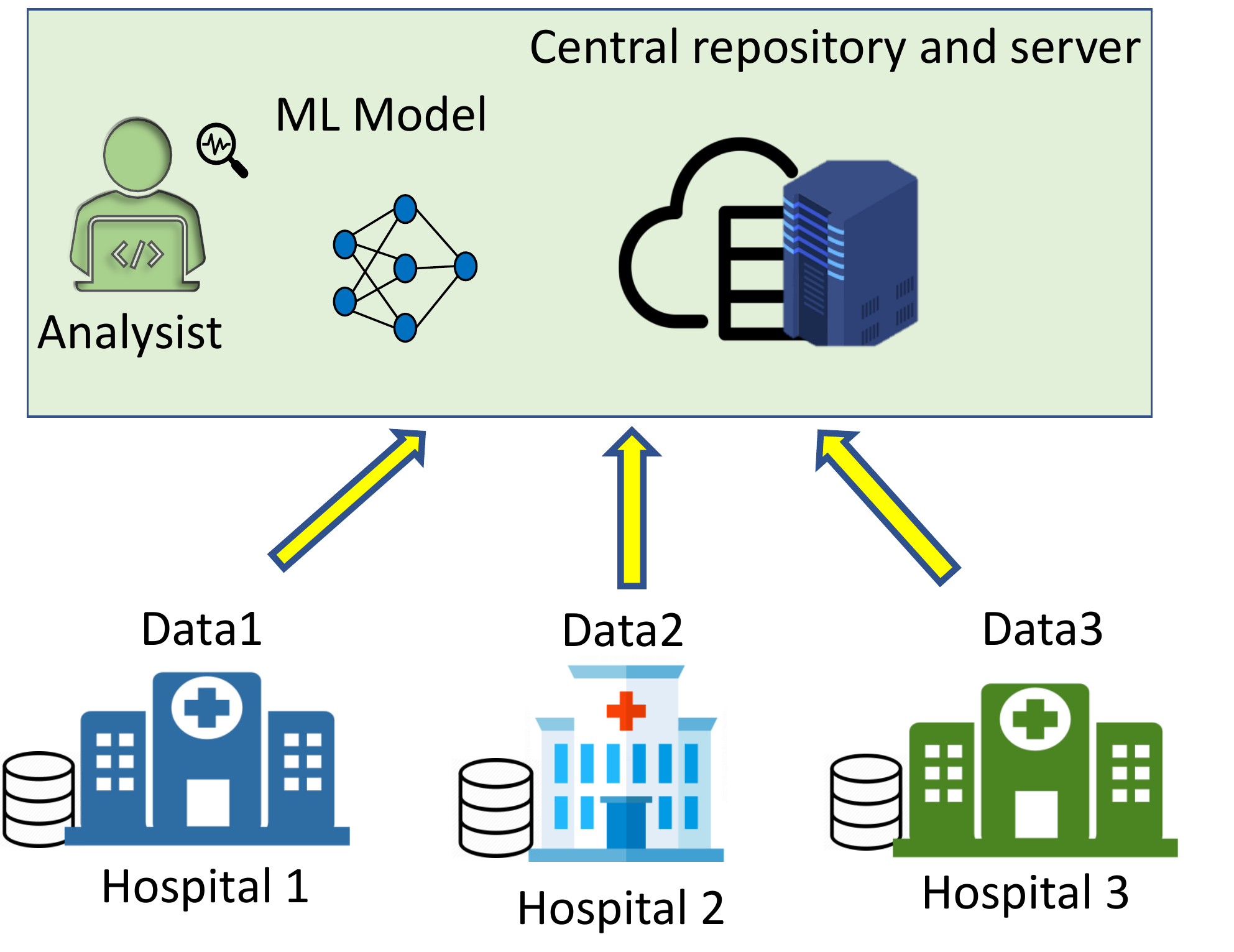}
			\label{fig:centrlaized_learning} 
		}
	\subfigure[]{
			\includegraphics[trim=10cm 0.5cm 2.5cm 0.1cm, clip=true, width=0.3\linewidth]{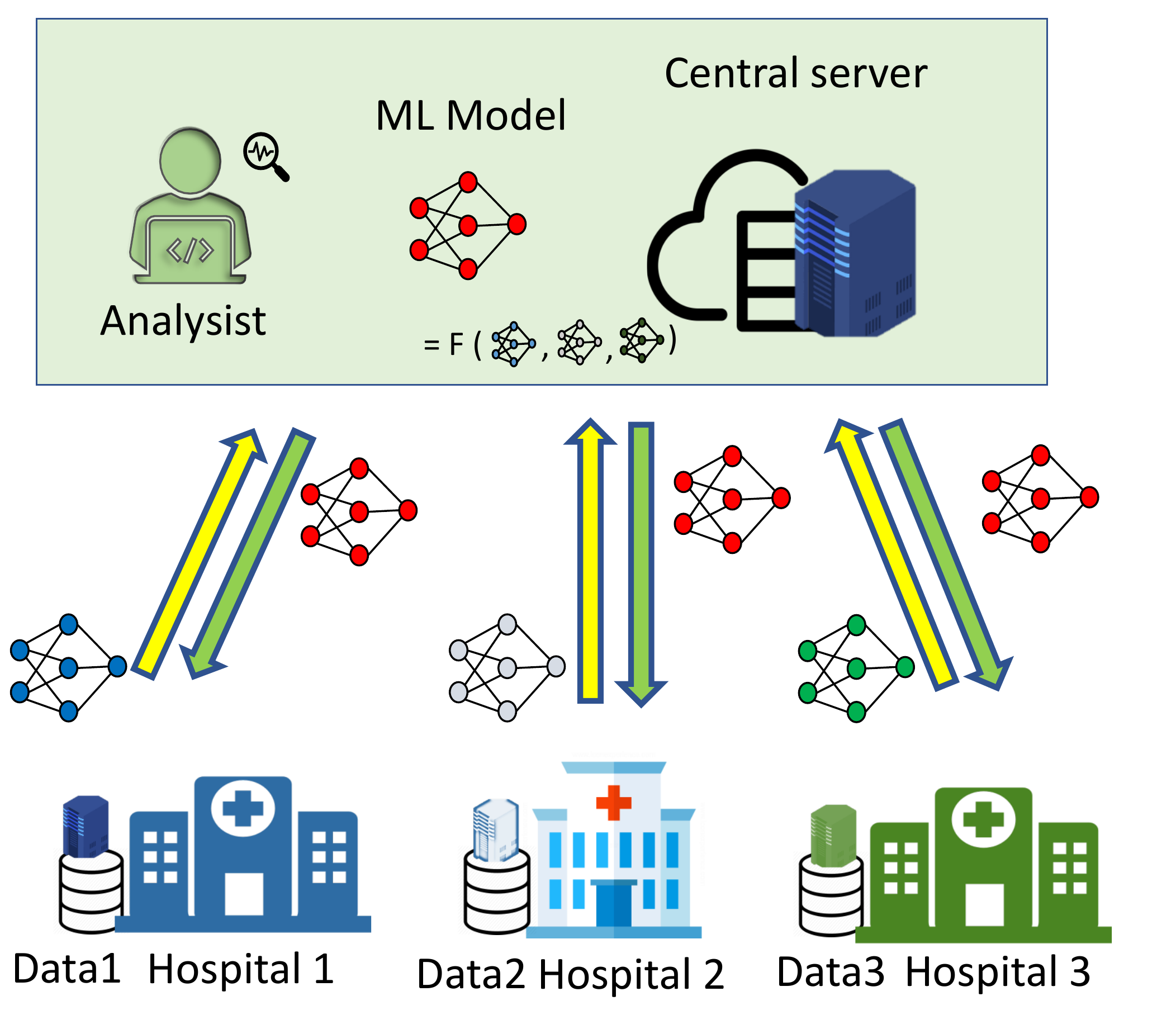}
			\label{fig:fed_learning_concept} 
		}
		\caption{\small{An illustration of (a) centralized learning (CL), and (b) federated learning (FL) with three hospitals. 
		}}
		\label{fig:flandcl}
\end{figure}

In centralized learning, data is collected to a central repository, and the analyst directly accesses those repositories to undertake machine learning training and inferences. In other words, this approach follows the data-to-model paradigm. 
This approach is not a privacy-preserving approach if the data is sensitive because analysts/researchers can directly access it. Fig.~\ref{fig:centrlaized_learning} illustrates centralized learning.

\label{sec:federated_learning}
Federated learning~\cite{fedkonecny2,fed1,reviewpaper_our} follows the model-to-data paradigm in which the model is sent to the clients (e.g., hospitals) to train/test instead of taking data out from those clients for the machine learning. 
Precisely, the model training/testing proceeds as follows: Let $W_t$ be the model parameters at time instance $t$ at the server. It is broadcast to all $N$ participating clients. Afterward, each client $i$ trains the model on their local data, and the locally trained model $W_t^{i}$, for $i \in \{1,2,\cdots,N\} $, is sent back to the server. Then the server aggregates the model, for example by weighted averaging, to form the global model $W_{t+1} = \sum_i \frac{n_i}{n}W_t^i$, where $n_i$ and $n$ are the number of samples at client $i$ and the total number of samples considering all clients. This process continues till the model converges. Fig.~\ref{fig:fed_learning} illustrates FL with three clients. As data always kept with the clients and the raw data is never seen by the analyst/server, FL provides privacy-by-design and privacy-by-default.

\section{Motivation of FedDICE for the ransomware spread detection}
\label{sec:fedDICE_motivation}

DICE enables collaboration among multiple hospitals to detect ransomware spread synergetically. In addition, FL enables all the DICE-specific requirements as listed in Section~\ref{sec:DICE}. This way, the integration of FL to DICE, which is forming FedDICE, is well motivated. Further, we elaborate on the two main reasons related to privacy and computation in the following. 

\subsection{Network flow dataset and its risk}
\label{sec:data-privacy}
Our dataset uses the network flow (netflow) format of CISCO. The fields include start time, source IP, Destination IP, total packets, total load, and source inter-packet arrival time~\cite{data_set}. Netflow is a protocol for collecting, aggregating, and recording traffic flow data in a network, gives you deep network visibility.

Netflow data of critical infrastructure such as ICE are sensitive. Adversaries can plan for an efficient attack in a network by gathering deep network information from the netflow data of the network. For example, (i) finding a bottleneck of the network can identify the possible target point for the attacker to craft a denial of service attack, and  (ii) finding the specific traffic patterns makes it easy for filtering attacks (blocking those traffic in the network). Consequently, a hospital does not want to share their netflow data with outsiders in DICE if raw ransomware data sharing is required. Works have been done to obfuscate the data in network flow~\cite{obfuscation}\cite{obfuscation2}. However, the sanitized data either decreases data quality for machine learning or poses some risk of leakage. The best option is not to share the (netflow) data in the first place but allow the machine learning inference. This is done by integrating FL in DICE, i.e., via FedDICE. Besides, FL enables privacy-by-design and privacy-by-default mechanisms, and it is compliant with data protection regulations such as GDPR~\cite{gdpr}.

\subsection{Distributed Computation}
The DICE formed by the collaboration of multiple hospitals (ICEs) is distributed and can be spread out to different geo-locations. DICE can leverage the advantages of distributed computing, including horizontal scaling of the network elements, fault tolerance, low latency, and distributed computational requirements. Being a distributed ML approach, FL enables all the benefits of the distributed computation to DICE. Thus FedDICE is promising for ransomware spread detection.


\section{Threat scenario} 
\label{sec:threat_scenario}
The threats in DICE can be ICE layer-specific, such as a compromised supervisor in the supervisory layer, malicious or infected devices, and malicious SDN network applications in the controller layer~\cite{alberto_work}. As the primary threat is usually due to the infected devices in an ICE, we consider only this threat scenario in this paper. In this scenario, firstly, a device is infected by a certain ransomware type. Then the ransomware spreads over the network to infect other devices/systems, including patient health monitoring devices and database servers, within the (local) ICE framework. 
During the development of the anti-ransomware model collaboratively in DICE, we assume that all participating hospitals are honest-but-curious parties. They behave as per our expectation but are only interested in inferring more on data of other hospitals. Also, they do not maliciously behave to alter or affect the anti-ransomware model training/testing process. We also assume that all the participating hospitals have the same network infrastructure and architecture in our studies for simplicity.

\section{Proposed security architecture of FedDICE}
\label{sec:proposed_solution}
\begin{figure}[t]
    \centering
    \includegraphics[width =0.5\linewidth]{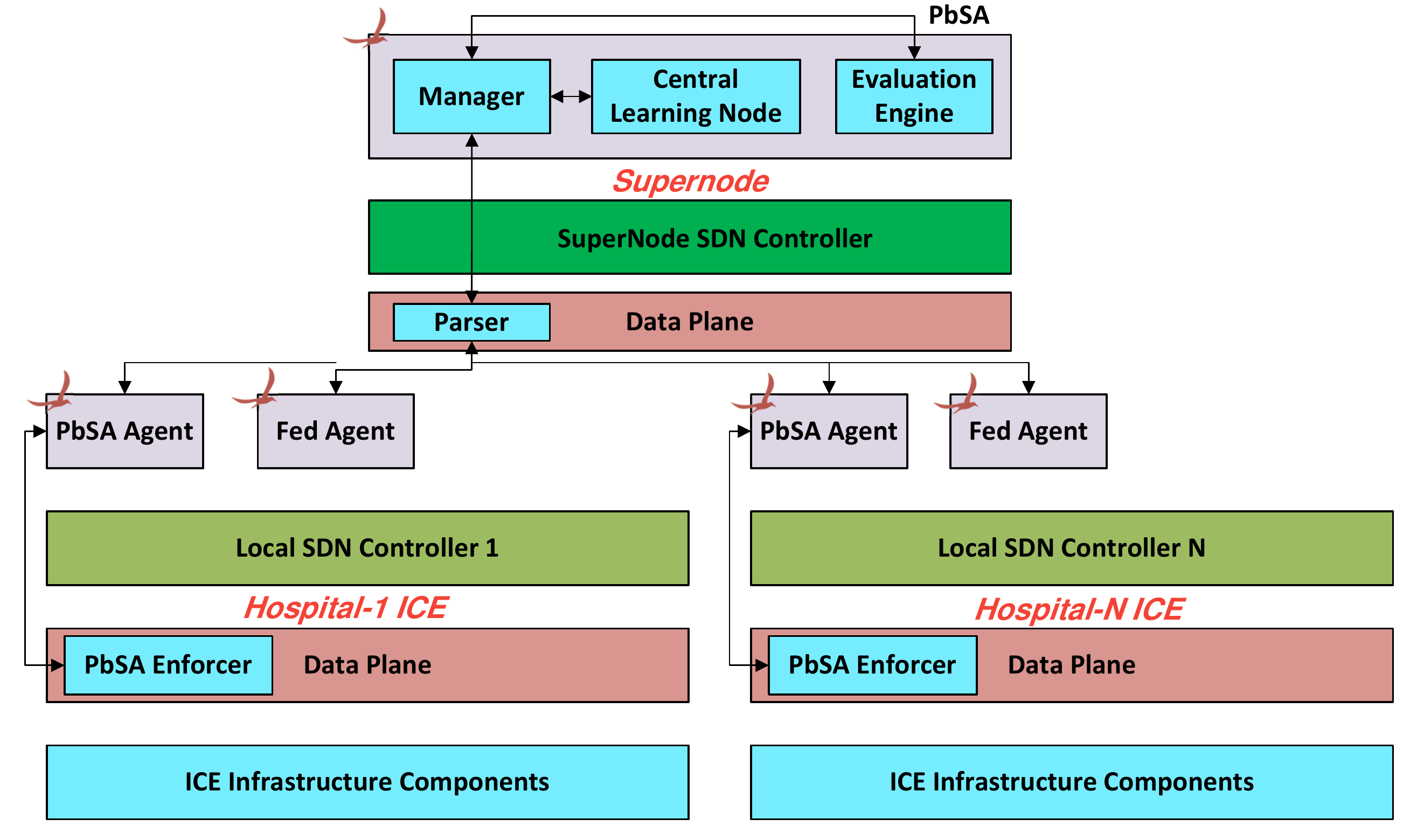}
    \caption{\small{Security architecture of FedDICE for the detection and mitigation of ransomware spread.}}
    \label{fig:secarc}
\end{figure}

Considering the threat scenario, we present a security architecture of FedDICE driven by SDN for the detection and mitigation of ransomware spread. This is an answer to \textbf{RQ2}. The architecture is depicted in Fig.~\ref{fig:secarc}, and it accounts for FedDICE with $N$ hospital ICEs. The major components and their functions in the architecture are presented in the following. 

\begin{enumerate}
    
    \item \textbf{Manager:} Manager is a software module present in the supernode. It has a collection of scripts that will be executed to manage FedDICE. The manager runs as a northbound application and interacts with the supernode SDN controller through its northbound application programming interface. It maintains communication with the local (leaf) SDN controller agent. Specifically, the manager performs the following tasks: (a) Maintains communication with the agents running on the local SDN controllers and the central learning node, (b) transforms the parser output to a more readable format for other software modules, (c) based on the central learning node feedback, the manager can create policies that can be installed in the policy repository maintained by the evaluation engine -- a system administrator can also install such policies in the policy repository -- and (d) conveys central learning node decisions to the enforcer module, and enforces appropriate flow rules in the data plane devices.

    \item \textbf{Evaluation Engine:} An evaluation engine is used to evaluate incoming network traffic against the relevant security policy expressions. Following the evaluation, the manager determines the flow rules, which are then conveyed to the enforcer module. This module also maintains a policy database. Our policy database is a simple JSON file; for example, refer to Fig.~\ref{fig:sample_policy}.

    \item \textbf{Agents:} Our security architecture uses agents, which are mainly northbound applications, to transfer information between the SDN Controllers. In some cases, they can be a separate application, which only utilises the controller information. This approach helps to build an application that can be used with a heterogeneous controller. The agents are running on each local SDN controller. Moreover, there are two agents; PbSA and Fed agents. PbSA is a policy-based security application running on the top of the local SDN controller. Firstly, it mirrors all the incoming packet information and formats them for machine learning purposes. This module also stores the formatted packet information in a local repository. The fed agent (federated machine agent) uses the local storage's raw data to build the local anti-malware model. Later it coordinates with the manager via parser to forward the local model and receive the global model. The global model is used to detect ransomware spread in local ICEs.

    \item \textbf{Enforcer:} We have modified the data plane devices and introduced an enforcer module. The enforcer module fetches the required information for the agents from the south-bound interface connected to the switches. It enables to enforcing flow rules obtained from the manager and applies the policies at the switching hardware.

    \item \textbf{Central Learning Node:} Central learning node is present in the supernode, and it is responsible for the aggregation of the anti-ransomware models received from the local ICEs during federated learning. The node also forms policies considering the aggregated models.

\end{enumerate}

Now we briefly discuss how the overall security components work together in FedDICE. Firstly, each local ICE fed agent regularly trains and updates its anti-ransomware ML model through FL. These models are used as policies in our approach. In contrast to a static policy, which is usually manually updated to detect and mitigate the ransomware spread, the ML model-based policy updates automatically through FL. Then it is applied to detect the ransomware spread in the local ICE. Afterward, the PbSA enforcer executes the policies to remove the infected devices from the ICE if there is any detection.


\section{Experiments, results, and analysis on the detection of ransomware spread}
\label{sec:dection}
In this section, we perform various experiments to investigate the performance of FedDICE in ransomware spread detection. This provides answers of \textbf{RQ1} and \textbf{RQ3}. Before presenting our results and analysis, we provide details on data, models, performance metrics, and implementation in the following sections.   

\subsection{Data}
We use a clinical environment network traffic dataset, where the data is captured in pcap and binetflow format~\cite{data_set}. 
The dataset has clean and ransomware-generated network traffic information by devices with an operating system, Windows (version: 7 and 10) and Ubuntu (version: 16.04). 
There are four popular ransomware families, namely \mbox{\emph{WannaCry}} \mbox{(RW-WC)}, \emph{Petya} (RW-PY), \emph{BadRabbit} (RW-BR) and \mbox{\emph{PowerGhost}} \mbox{(RW-PG)}. 
The dataset is generated by considering sliding windows of netflows to obtain a total of 520 features related to transmission control protocol (TCP), user datagram protocol (UDP), address resolution protocol (ARP), and flow features. The sliding windows can have various time duration, including 5s, 10s, and 20s. In the feature generation process, aggregated features are generated using the last flow information of the sliding window duration (e.g., 10s). This is done to collect a wide variety of both the context features and flow features. Refer to~\cite{sensor_paper} for details.    
The samples refer to the aggregated traffic information at different time frames. There are 150540 samples if we use a 10s window in our network observations. 
For all our experiments, we use the dataset split and numbers as depicted in Table~\ref{tab:1}.

\begin{table}[tbh]
\centering
\scriptsize
\vspace{-15pt}
\caption{\small{Dataset (prepared by using 10s sliding window duration)}}
\begin{tabular}{ccccc}
\toprule
Dataset type & Train  & Validataion & Test  & Total  \\ \midrule \midrule
Clean        & 80000  & 10000       & 10000 & 100000 \\
RW-WC     & 20000  & 2500        & 2500  & 25000  \\
RW-PY        & 784    & 98          & 99    & 981    \\
RW-BR    & 311    & 38          & 40    & 389    \\
RW-PG   & 19336  & 2417        & 2417  & 24170  \\ \midrule
\cellcolor[HTML]{9AFF99} Total        & \cellcolor[HTML]{9AFF99} 120431 & \cellcolor[HTML]{9AFF99} 15053       & \cellcolor[HTML]{9AFF99} 15056 & \cellcolor[HTML]{9AFF99} 150540 \\ \bottomrule
\end{tabular}
\label{tab:1}
\end{table}

We divide the dataset into two categories during training: clean (label 1) and ransomware (label 0). We consider binary classification because our main task is to detect the ransomware irrespective of its specific type. 
In FL, the training and validation dataset are local, so it is different in different clients, whereas the test dataset is global and checked on the global model. 

\subsection{Model}
\label{sec:model}
We consider three different models commonly used in ransomware detection and compatible for FL in our analyses. All are considered for binary classification tasks with an observation window of 10 global epochs in FL.
\begin{enumerate} 
    \item \textbf{Logistic regression}: Assume $W$ and $b$ be the weight and bias of the model, respectively. The output of the logistic regression (LR) model is given by: $\sigma( \mathbf{w}^{\textup{T}} \mathbf{x} + b)$, where $\sigma$ is the sigmoid function, $\mathbf{w}$ is the parameter vector, and $\mathbf{x}$ is the input. The sigmoid function maps any real input to the value between zero and one. 
    We consider a binary logistic regression with cross-entropy loss and learning rate = 0.01 in our analyses. 
    
    \item \textbf{Support vector machine}: Support vector machine (SVM) uses the kernel trick (data is projected to higher dimensions) to express its linear function by: $b + \sum_i \alpha_i k (\mathbf{x},\mathbf{x}^{(i)})$, where $k (\mathbf{x},\mathbf{x}^{(i)}) = \phi(x). \phi(x^{(i)}$ is called a kernel, $\mathbf{x}^{(i)}$ is a training example, and $\alpha$ is a coefficient vector. The output function is non-linear to $x$ in SVM.   
    We consider the SVM model for binary classification (i.e., C=1) in our analyses. We utilize the skitlearn library to implement the SVM. We use the default parameters (e.g., kernel = rbf, degree = 3, gamma = scale, and max\_iter = -1).
    
    \item \textbf{Feedforward Neural Network}: Feedforward neural network (FNN) is formed by the composition of multiple neural network layers, where each layer performs some functions in its input as follows: The first layer output is $\mathbf{h}^{(1)} = g^{(1)} (\mathbf{w}^{ (1)\textup{T}} \mathbf{x} +b^{(1)})$, the second layer output is $\mathbf{h}^{(2)} = g^{(2)} (\mathbf{w}^{(2)\textup{T}} \mathbf{h}^{\textup{1}} +b^{(2)})$, and similar calculations proceeds till the output layer. The layers between the input and output are called hidden layers. In our analyses, we consider a FNN with six hidden layers (each followed by batch-normalization and Relu activation) and softmax output layer. (num features, 1024), (1024, 512), (512, 128), (128, 64), (64, 32), and (32, num class) are the pairs of the number of nodes in hidden layers starting from first to the last, respectively. The cross-entropy loss and learning rate equals 0.01 are used in training. 
\end{enumerate}

\subsection{Metrics for performance measurement}
We consider accuracy, precision, recall, F1-score, and false-negative rate (FNR) for performance analysis. These metrics are defined in Appendix~\ref{appendix:metrics}. In our dataset setup, the ransomware class is labeled 0, and it is a positive class, whereas the normal class is labeled 1, and it is a negative class. Thus, we consider only FNR (not false-positive rate) in our analyses. Moreover, precision, recall, and F1-score are averaged figures (average = `macro'). Besides, 4 fold validation is carried out in all of our experiments. 

\subsection{Implementation}   
The training/testing implementation is done in python programming by leveraging PyTorch library~\cite{pytorch}. Moreover, for FL with logistic regression and SVM models, we have used some functions from Sherpa.ai Federated Learning and Differential Privacy Framework~\cite{sherpaai}. All the programs were executed on a Dell laptop with Intel Core i5-8350U CPU, 8GB RAM, and x64-based processor.
%

\subsection{Results and analysis}
\label{sec:results_and_analysis}
We present our results and analyze them on the detection of ransomware spread in this section. Firstly, our baseline performance provided by centralized learning is presented, and then we proceed to find the FL performances under three ML models defined in section~\ref{sec:model}.  

\subsubsection{(Baseline performance) Centralized learning:}
\begin{table}[thb]
\centering
\scriptsize
\caption{\small{Results summary for centralized learning on testing dataset.}}
\vskip2pt
\begin{tabular}{@{}cccccc@{}}
\toprule
\rowcolor[HTML]{FFFFFF} 
Model & Accuracy & Precision & Recall & F1-score & FNR \\ \midrule
\rowcolor[HTML]{9AFF99} 
LR & 0.999768 & 0.99975 & 0.999733 & 0.999738 & 0.000395 \\ \midrule
SVM & 0.999395 & 0.999538 & 0.999105 & 0.999323 & 1.78E-03 \\
FNN & 0.98954 & 0.992003 & 0.98464 & 0.988185 & 0.030271 \\ \bottomrule
\end{tabular}%
\label{tab:2}
\end{table}

\hskip2pt Our results on the test dataset (in Table~\ref{tab:2}) and validation dataset (in Appendix~\ref{appendix:val_results}) showed excellent similar performance by all three models in centralized learning. For example, we get around 99.9\% accuracy for LR and SVM and around 99\% for FNN. In addition, all models return almost zero FNR. Relatively, logistic regression is slightly better than others.  
%

\subsubsection{(RQ1) Need of a collaborative framework in ransomware spread detection:}

\hskip2pt Under \textbf{RQ1}, we are aiming to find if a model trained on one ransomware family be still effective in detecting another ransomware family. If the answer is negative in general, then the single client's model and data are not sufficient to capture unseen ransomware characteristics. This way, we demonstrate the need for collaborative machine learning such as FL in DICE for the detection. 

Now considering each client (i.e., local ICE) is infected by only one ransomware family, we form an experimental setup of the data distribution as depicted in Table~\ref{tab:data_distribution}. Starting from Client1, a model is trained on its training data (consisting of clean and RW-WC), and it is tested against the training data of clients 2, 3, and 4 (that have clean and other types of ransomware samples). Similarly, we carried experiments for all clients. The results for FR, SVM and FNN are presented in Table~\ref{tab:log_results}, \ref{tab:SVM_results}, and \ref{tab:NN_results}, respectively.  

\begin{table}[tbh]
\centering
\caption{\small{Data distribution among clients.}}
\vskip2pt
\resizebox{0.5\textwidth}{!}{%
\begin{tabular}{@{}c |c |c | c | c@{}}
\toprule
Clients & Client1 & Client2 & Client3 & Client4 \\ \midrule \midrule
\begin{tabular}[c]{@{}c@{}}Training Data\\ and Samples\end{tabular} & \begin{tabular}[c]{@{}c@{}}Clean (20000)\\ RW-WC (20000)\end{tabular} & \begin{tabular}[c]{@{}c@{}}Clean (20000)   \\ RW-PY (784)\end{tabular} & \begin{tabular}[c]{@{}c@{}}Clean (20000)\\ RW-BR (310)\end{tabular} & \begin{tabular}[c]{@{}c@{}}Clean (20000)\\ RW-PG (19336)\end{tabular} \\ \midrule
\begin{tabular}[c]{@{}c@{}} Training\\ Sample Size \end{tabular} & 40,000 & 20,784 & 20,310 & 39,336 \\
\midrule \midrule
\multicolumn{5}{l}{Test dataset has 20000 Clean, 5000 RW-WC, 197 RW-PY, 79 RW-BR and 4834 RW-PG samples}\\
\bottomrule
\end{tabular}%
}
\label{tab:data_distribution}
\end{table}

\begin{table}[tbh]
\centering
\scriptsize
\vskip-10pt
\caption{\small{(\textbf{RQ1}) Results for the logistic regression (LR) when training on only one client and testing on the remaining clients for the setup in Table~\ref{tab:data_distribution}.}}
\vskip2pt
\resizebox{0.5\textwidth}{!}{%
\begin{tabular}{@{}|c|c|c|c|c|c|c|c|@{}}
\toprule
Training & Testing & Accuracy & Precision & Recall & F1-score & FNR & \begin{tabular}[c]{@{}c@{}}Misclassified \\ ransomware \\ samples\end{tabular} \\ \midrule
 & Client2 & 0.96228 & 0.98114 & 0.5 & 0.49039 & \cellcolor[HTML]{9AFF99}1 & \cellcolor[HTML]{9AFF99}784 out of 784 \\ \cmidrule(l){2-8} 
 & Client3 & 0.98474 & 0.99237 & 0.5 & 0.49615 & \cellcolor[HTML]{9AFF99}1 & \cellcolor[HTML]{9AFF99}310 out of 310 \\ \cmidrule(l){2-8} 
 & Client4 & 0.99497 & 0.9951 & 0.99488 & 0.99496 & 0.01024 & 198 out of 19336 \\ \cmidrule(l){2-8} 
\multirow{-4}{*}{Client1} & Test dataset & 0.99083 & 0.99319 & 0.98635 & 0.98965 & 0.0273 & 276 out of 10110 \\ \midrule \midrule
 & Client1 & 0.93015 & 0.93865 & 0.93015 & 0.92981 & 0.13945 & 2789 out of 20000 \\ \cmidrule(l){2-8} 
 & Client3 & 0.9997 & 0.99665 & 0.9935 & 0.99507 & 0.012903 & 4 out of 310 \\ \cmidrule(l){2-8} 
 & Client4 & 0.99939 & 0.99939 & 0.99939 & 0.99939 & 0.000879 & 17 out of 19336 \\ \cmidrule(l){2-8} 
\multirow{-4}{*}{Client2} & Test dataset & 0.92016 & 0.94635 & 0.88111 & 0.90419 & 0.237784 & 2404 out of 10110\\ \midrule \midrule
 & Client1 & 0.50255 & 0.74583 & 0.50255 & 0.33902 & \cellcolor[HTML]{9AFF99}0.99485 & \cellcolor[HTML]{9AFF99}19897 out of 20000 \\ \cmidrule(l){2-8} 
 & Client2 & 0.99682 & 0.99836 & 0.95791 & 0.97721 & 0.084184 & 66 out of 784 \\ \cmidrule(l){2-8} 
 & Client4 & 0.52433 & 0.75832 & 0.51616 & 0.37196 & \cellcolor[HTML]{9AFF99}0.967677 & \cellcolor[HTML]{9AFF99}18711 out of 19336 \\ \cmidrule(l){2-8} 
\multirow{-4}{*}{Client3} & Test dataset & 0.6716 & 0.83458 & 0.51098 & 0.42238 & \cellcolor[HTML]{9AFF99}0.978042 & \cellcolor[HTML]{9AFF99}9888 out of 10110 \\ \midrule \midrule
 & Client1 & 1 & 1 & 1 & 1 & 0 & 0 out of 20000 \\ \cmidrule(l){2-8} 
 & Client2 & 0.97638 & 0.98802 & 0.68686 & 0.76599 & \cellcolor[HTML]{9AFF99}0.626276 & \cellcolor[HTML]{9AFF99}491 out of 784 \\ \cmidrule(l){2-8} 
 & Client3 & 0.99301 & 0.99648 & 0.77097 & 0.8497 & \cellcolor[HTML]{9AFF99}0.458065 & \cellcolor[HTML]{9AFF99}142 out of 310 \\ \cmidrule(l){2-8} 
\multirow{-4}{*}{Client4} & Test dataset & 0.99415 & 0.99564 & 0.9913 & 0.99342 & 0.017409 & 176 out of 10110\\ \bottomrule
\end{tabular}%
}
\label{tab:log_results}
\vskip-12pt
\end{table}
\begin{table}[tbh]
\centering
\scriptsize
\caption{\small{(\textbf{RQ1}) Results for the SVM when training on only one client and testing on the remaining clients for the setup in Table~\ref{tab:data_distribution}.}}
\vskip2pt
\resizebox{0.5\textwidth}{!}{%
\begin{tabular}{@{}|c|c|c|c|c|c|c|c|@{}}
\toprule
Training & Testing & Accuracy & Precision & Recall & F1-score & FNR & \begin{tabular}[c]{@{}c@{}}Misclassified \\ ransomware \\ samples\end{tabular} \\ \midrule
 & Client2 & 0.96228 & 0.98114 & 0.5 & 0.49039 & \cellcolor[HTML]{9AFF99}1 & \cellcolor[HTML]{9AFF99}784 out of 784 \\ \cmidrule(l){2-8} 
 & Client3 & 0.98474 & 0.99237 & 0.5 & 0.49615 & \cellcolor[HTML]{9AFF99}1 & \cellcolor[HTML]{9AFF99}310 out of 310 \\ \cmidrule(l){2-8} 
 & Client4 & 0.9912 & 0.9915 & 0.99105 & 0.9912 & 0.017894 & 346 out of 19336 \\ \cmidrule(l){2-8} 
\multirow{-4}{*}{Client1} & Test dataset & 0.99083 & 0.99319 & 0.98635 & 0.98965 & 0.0273 & 276 out of 10110 \\ \midrule \midrule
 & Client1 & 0.5022 & 0.73459 & 0.5022 & 0.33834 & \cellcolor[HTML]{9AFF99}0.99545 & \cellcolor[HTML]{9AFF99}19909  out of 20000 \\ \cmidrule(l){2-8} 
 & Client3 & 0.99936 & 0.9964 & 0.98221 & 0.9892 & 0.035484 & 11 out of 310 \\ \cmidrule(l){2-8} 
 & Client4 & 0.51993 & 0.7539 & 0.51169 & 0.36262 & \cellcolor[HTML]{9AFF99}0.976469 & \cellcolor[HTML]{9AFF99}18881 out of 19336 \\ \cmidrule(l){2-8} 
\multirow{-4}{*}{Client2} & Test dataset & 0.67164 & 0.83459 & 0.51103 & 0.42249 & \cellcolor[HTML]{9AFF99}0.977943 & 9887 out of 10110 \\ \midrule \midrule
 & Client1 & 0.5014 & 0.75035 & 0.5014 & 0.33644 & \cellcolor[HTML]{9AFF99}0.9972 & \cellcolor[HTML]{9AFF99}19944 out of 20000 \\ \cmidrule(l){2-8} 
 & Client2 & 0.99601 & 0.99793 & 0.94707 & 0.97102 & 0.105867 & 83 out of 784 \\ \cmidrule(l){2-8} 
 & Client4 & 0.51792 & 0.75665 & 0.50965 & 0.35812 & \cellcolor[HTML]{9AFF99}0.98071 & \cellcolor[HTML]{9AFF99}18963 out of 19336 \\ \cmidrule(l){2-8} 
\multirow{-4}{*}{Client3} & Test dataset & 0.6715 & 0.83455 & 0.51083 & 0.42208 & \cellcolor[HTML]{9AFF99}0.978338 & 9891 out of 10110 \\ \midrule \midrule
 & Client1 & 1 & 1 & 1 & 1 & 0 & 0 out of 20000 \\ \cmidrule(l){2-8} 
 & Client2 & 0.96339 & 0.98167 & 0.51467 & 0.51917 & \cellcolor[HTML]{9AFF99}0.970663 & \cellcolor[HTML]{9AFF99}761 out of 784 \\ \cmidrule(l){2-8} 
 & Client3 & 0.98567 & 0.99283 & 0.53065 & 0.55414 & \cellcolor[HTML]{9AFF99}0.93871 & \cellcolor[HTML]{9AFF99}291 out of 310 \\ \cmidrule(l){2-8} 
\multirow{-4}{*}{Client4} & Test dataset & 0.99117 & 0.99344 & 0.98684 & 0.99003 & 0.026311 & 266 out of 10110 \\ \bottomrule
\end{tabular}%
}
\label{tab:SVM_results}
\vskip-12pt
\end{table}
\begin{table}[tbh]
\centering
\scriptsize
\caption{\small{(\textbf{RQ1}) Results for the Feedforward Neural Network (FNN) when training on only one client and testing on the remaining clients for the setup in Table~\ref{tab:data_distribution}.}}
\vskip2pt
\resizebox{0.5\textwidth}{!}{%
\begin{tabular}{@{} |c|c|c|c|c|c|c|c| @{}}
\toprule
Training & Testing & Accuracy & Precision & Recall & F1-score & FNR & \begin{tabular}[c]{@{}c@{}}Misclassified \\ ransomware \\ samples\end{tabular} \\ \midrule
 & Client2 & 0.96093 & 0.55625 & 0.50298 & 0.49732 & \cellcolor[HTML]{9AFF99}0.992347 & \cellcolor[HTML]{9AFF99}778 out of 784 \\ \cmidrule(l){2-8} 
 & Client3 & 0.98336 & 0.5675 & 0.50883 & 0.51295 & \cellcolor[HTML]{9AFF99}0.980645 & \cellcolor[HTML]{9AFF99}304 out of 310 \\ \cmidrule(l){2-8} 
 & Client4 & 0.97272 & 0.97438 & 0.97228 & 0.97267 & 0.054096 & 1046 out of 19336 \\ \cmidrule(l){2-8} 
\multirow{-4}{*}{Client1} & Test dataset & 0.9903 & 0.99232 & 0.986 & 0.98906 & 0.027102 & 276 out of 10110 \\ \midrule \midrule
 & Client1 & 0.5803 & 0.66871 & 0.5803 & 0.51703 & \cellcolor[HTML]{9AFF99}0.78165 & \cellcolor[HTML]{9AFF99}15633  out of 20000 \\ \cmidrule(l){2-8} 
 & Client3 & 0.94353 & 0.59966 & 0.92528 & 0.64968 & 0.093548 & 29 out of 310 \\ \cmidrule(l){2-8} 
 & Client4 & 0.64935 & 0.72344 & 0.64437 & 0.61314 & \cellcolor[HTML]{9AFF99}0.650703 & \cellcolor[HTML]{9AFF99}12582 out of 19336 \\ \cmidrule(l){2-8} 
\multirow{-4}{*}{Client2} & Test dataset & 0.79389 & 0.80031 & 0.72201 & 0.73978 & \cellcolor[HTML]{9AFF99}0.496835 & \cellcolor[HTML]{9AFF99} 378 out of 10110 \\ \midrule \midrule
 & Client1 & 0.74245 & 0.75772 & 0.74245 & 0.73858 & \cellcolor[HTML]{FFFFFF}0.37925 & \cellcolor[HTML]{FFFFFF}7585 out of 20000 \\ \cmidrule(l){2-8} 
 & Client2 & 0.84729 & 0.52867 & 0.59466 & 0.52663 & \cellcolor[HTML]{9AFF99}0.678571 & \cellcolor[HTML]{9AFF99}532 out of 784 \\ \cmidrule(l){2-8} 
 & Client4 & 0.76668 & 0.77746 & 0.76496 & 0.76355 & \cellcolor[HTML]{FFFFFF}0.33704 & \cellcolor[HTML]{FFFFFF}6517 out of 19336 \\ \cmidrule(l){2-8} 
\multirow{-4}{*}{Client3} & Test dataset & 0.76008 & 0.73263 & 0.7116 & 0.71925 & \cellcolor[HTML]{9AFF99}0.436004 & \cellcolor[HTML]{9AFF99} 9860 out of 10110 \\ \midrule \midrule
 & Client1 & 0.99292 & 0.99302 & 0.99293 & 0.99292 & 0 & 0 out of 20000 \\ \cmidrule(l){2-8} 
 & Client2 & 0.96276 & 0.74198 & 0.65957 & 0.69132 & \cellcolor[HTML]{9AFF99}0.668367 & \cellcolor[HTML]{9AFF99}524 out of 784 \\ \cmidrule(l){2-8} 
 & Client3 & 0.98124 & 0.70533 & 0.78246 & 0.73744 & \cellcolor[HTML]{9AFF99}0.422581 & \cellcolor[HTML]{9AFF99}131 out of 310 \\ \cmidrule(l){2-8} 
\multirow{-4}{*}{Client4} & Test dataset & 0.98482 & 0.98291 & 0.98307 & 0.98299 & 0.022255 & 752 out of 10110\\ \bottomrule
\end{tabular}%
}
\label{tab:NN_results}
\vskip-8pt
\end{table}
We summarize our observations based on the empirical results depicted in Tables~\ref{tab:log_results}, \ref{tab:SVM_results}, and \ref{tab:NN_results} in the following:
\begin{enumerate} 
    \item For all our models trained on WannaCry (RW-WC) and clean dataset (i.e., Client1) has shown good performance on PowerGhost (RW-PG) and test dataset, but detect only a few samples of Petya (RW-PY) and BadRabbit (RW-BR). This is indicated by a high FNR greater than 0.98 for the cases.
    \item For all our models trained on RW-PG and clean dataset (i.e., Client4) has shown good performance on RW-WC and test dataset. There is a high FNR for RW-PY and RW-BR. 
    \item For all our models trained on RW-PY and clean dataset (i.e., Client2) has shown good performance only on RW-BR (i.e., Client3), and vice-versa. 
\end{enumerate}
These results show that there can be some ransomware such as RW-WC and RW-PG, those indicating a possibility of using models trained on network flow information generated by one ransomware to detect each other. However, this is not true in general; for example, models trained on RW-WC or RW-PG have poor performance on RW-PY and RW-BR. This clearly shows that we need collaboration either by direct data sharing or using privacy-preserving techniques such as FL in DICE. The results from the direct data-sharing approach are equivalent to the centralized learning, so we further analyze FL in the following section. 

\subsubsection{(RQ3) Performance of FedDICE over centralized learning:}
\hskip2pt Under \textbf{RQ3}, FedDICE, specifically FL, is investigated considering setups with three and four clients, separately. The number of clients is chosen based on possible data distribution and the number of ransomware types. As the FL's performance is dependent on the type of data distribution, we consider both the IID and the non-IID data (representing the heterogeneity of data) distribution among the clients as described in the following:
\begin{enumerate} 
    \item (\textbf{CASE-I}) IID data distribution: Each client has clean and all samples of ransomware that are independently and identically distributed. The total sample size in each client is the same. We call \textbf{CASE I-A} and \textbf{CASE I-B} for three clients and four client setups, respectively. 
    \item (\textbf{CASE-II}) Non-IID data distribution: The distribution of the non-IID datasets is due to the label distribution skew. If three clients, we say \textbf{CASE II-A}, where Client1 has normal and RW-WC samples, Client2 has normal, RW-PY and RW-BR samples, and Client3 has normal and RW-PG samples. Another case is with the four clients, we say \textbf{CASE II-B}, and its data distribution is depicted in Table~\ref{tab:data_distribution}.
\end{enumerate}


\begin{table}[t]
\centering
\scriptsize
\vskip-10pt
\caption{\small{ (\textbf{RQ3}) Results summary for the global model over the testing dataset in federated learning and (a) \textbf{CASE I-A}, and (b) \textbf{CASE I-B}.}}
\resizebox{0.8\textwidth}{!}{
	\subfigure[]{
    	\begin{tabular}{@{}cccccc@{}}
            \toprule
            \rowcolor[HTML]{FFFFFF} 
            Model & Accuracy & Precision & Recall & F1-score & FNR \\ \midrule
            \rowcolor[HTML]{9AFF99} 
            LR & 0.999725 & 0.999728 & 0.999658 & 0.99969 & 0.000544 \\ \midrule
            SVM & 0.999453 & 0.999578 & 0.999198 & 0.999385 & 0.001583 \\ \midrule
            FNN & 0.99881 & 0.999073 & 0.998263 & 0.998665 & 0.003401 \\ \bottomrule
        \end{tabular}
        \label{tab:caseIA}
	}
	\hskip20pt
		\subfigure[]{
		\begin{tabular}{@{}cccccc@{}}
            \toprule
            \rowcolor[HTML]{FFFFFF} 
            Model & Accuracy & Precision & Recall & F1-score & FNR \\ \midrule
            \rowcolor[HTML]{9AFF99} 
            LR & 0.999675 & 0.999708 & 0.999565 & 0.999635 & 0.000767 \\ \midrule
            SVM & 0.999428 & 0.999558 & 0.999163 & 0.999358 & 0.001656 \\ \midrule
            FNN & 0.998188 & 0.99854 & 0.99739 & 0.99796 & 0.005007 \\ \bottomrule
        \end{tabular}
        \label{tab:caseIB}
		}}
		\vskip-10pt
\end{table}

\begin{figure}[tbh]
	\centering
		\subfigure[]{
			\includegraphics[width=0.48\linewidth]{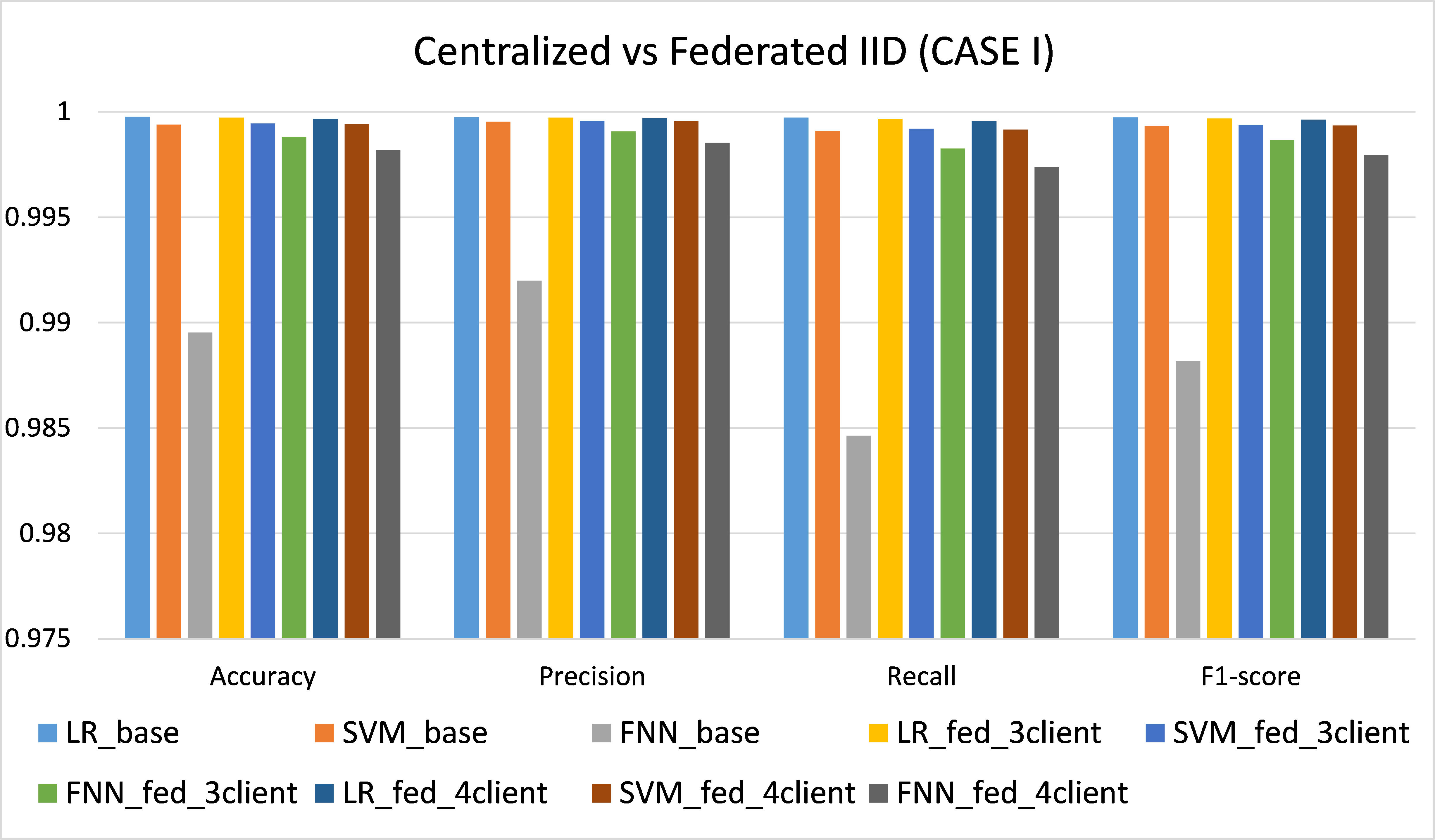}
		}
		\hskip-10pt
		\subfigure[]{
			\includegraphics[width=0.48\linewidth]{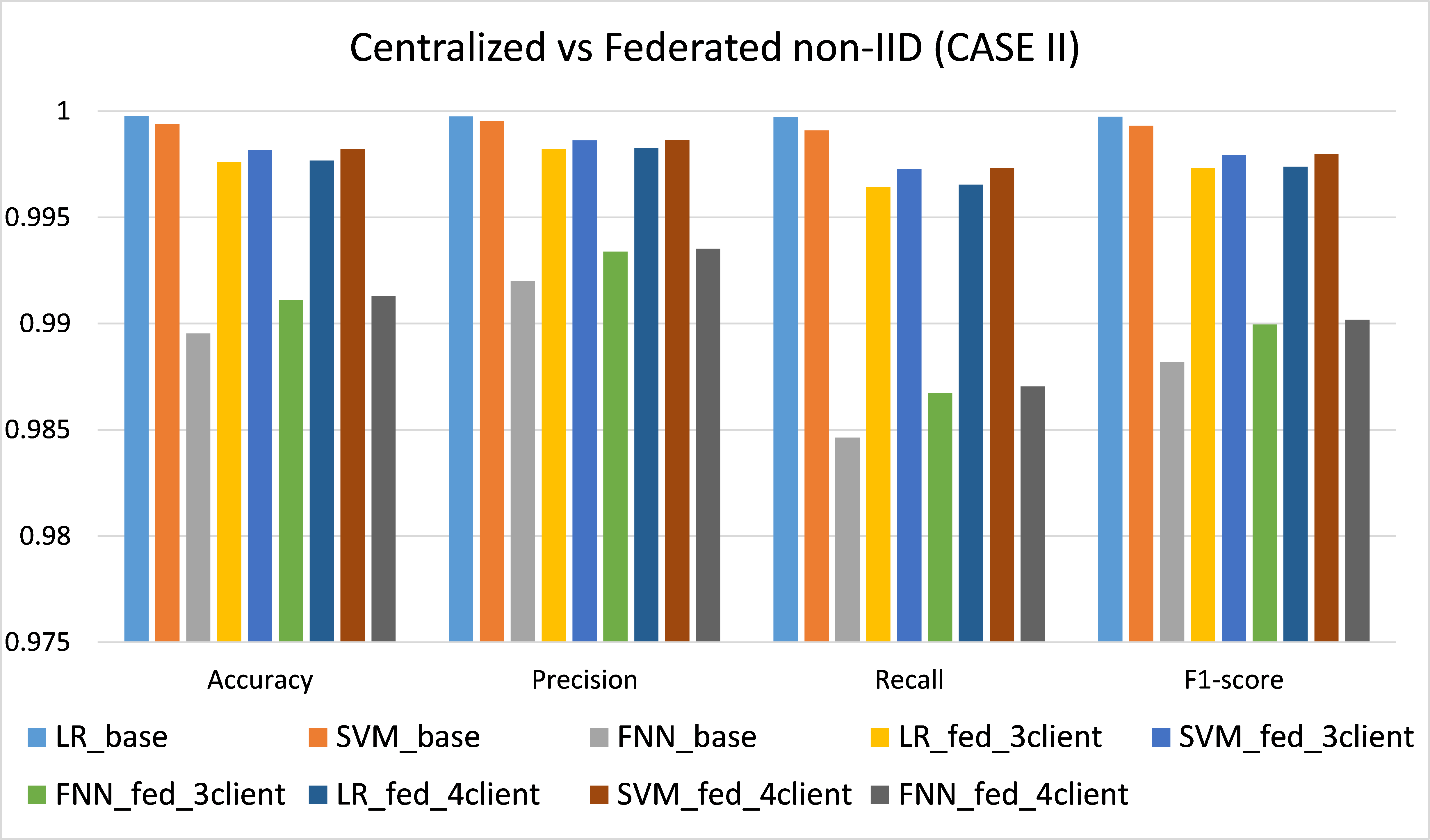}
		}
		\vskip-6pt
		\caption{\small{Performances of the centralized and federated learning (a) \textbf{CASE-I}  and (b) \textbf{CASE-II} in bar diagrams where the y-axis's units are shown in the interval of 0.005 units. In labels, the word ``\emph{base}" refers to the centralized learning, and ``\emph{fed}'' refers to the federated learning.}}
		\label{fig:combined_results}
		\vskip5pt
	\end{figure}

For the \textbf{CASE-I}, results (in Table~\ref{tab:caseIA} and \ref{tab:caseIB}) show that FL is effective in ransomware spread detection over the IID data distribution across participating clients. Moreover, its performance is similar to centralized learning (refer to Fig.~\ref{fig:combined_results}(a)), and all three models achieve the best performance; however, the logistic regression (LR) has a relatively slightly better performance. Usually, FL performance is lower than centralized learning (baseline), but we observe a difference in our results. For example, the performance of FNN in FL is higher than its baseline (see Fig.~\ref{fig:combined_results} (a)). The possible reason for this result can be the effect of model aggregation in FL. The weighted averaging of the models in model aggregation can stabilize the fluctuations in the model updates to some extent during the model training in FNN and thus contribute to better performance. 

\begin{table} [t]
\centering
\scriptsize
\caption{\small{ (\textbf{RQ3}) Results summary for the global model over the testing dataset in federated learning and (a) \textbf{CASE II-A}, and (b) \textbf{CASE II-B}.}}
\resizebox{0.8\textwidth}{!}{
	\subfigure[]{
    	\begin{tabular}{@{}cccccc@{}}
        \toprule
        \rowcolor[HTML]{FFFFFF} 
        Model & Accuracy & Precision & Recall & F1-score & FNR \\ \midrule
        \rowcolor[HTML]{FFFFFF} 
        LR & 0.99761 & 0.99821 & 0.99644 & 0.99731 & 0.007122 \\ \midrule
        \rowcolor[HTML]{9AFF99} 
        SVM & 0.99817 & 0.99863 & 0.99728 & 0.99795 & 0.00544 \\ \midrule
        FNN & 0.9911 & 0.99339 & 0.98675 & 0.98996 & 0.026508 \\ \bottomrule
        \end{tabular}%
        \label{tab:caseIIA}
	}
	\hskip20pt
		\subfigure[]{
		        \begin{tabular}{@{}cccccc@{}}
                \toprule
                \rowcolor[HTML]{FFFFFF} 
                Model & Accuracy & Precision & Recall & F1-score & FNR \\ \midrule
                \rowcolor[HTML]{FFFFFF} 
                LR & 0.99768 & 0.99826 & 0.99654 & 0.99739 & 0.006924 \\ \midrule
                \rowcolor[HTML]{9AFF99} 
                SVM & 0.99821 & 0.99865 & 0.99733 & 0.99799 & 0.005341 \\ \midrule
                FNN & 0.9913 & 0.99353 & 0.98704 & 0.99018 & 0.025915 \\ \bottomrule
                \end{tabular}%
                \label{tab:caseIIB}
		}}
		\vskip-10pt
\end{table}



For the \textbf{CASE-II}, results (in Table~\ref{tab:caseIIA} and \ref{tab:caseIIB}) show that FL is effective in ransomware spread detection even over the non-IID data distribution across participating clients. Moreover, like in the IID case, its performance is similar to centralized learning (refer to Fig.~\ref{fig:combined_results}(b)), and all three models achieve the best performance; however, the SVM model has relatively slightly better performance. As expected, due to the skewness in the label distribution over the clients in non-IID data distribution across the participating clients, the FL performance is relatively lower than its IID counterparts. However, the overall maximum degradation is only on a scale of around 2 digits after the decimal, which is insignificant (e.g., 0.01).    

\textbf{Remark:} Based on the overall results obtained with FL in CASE-I and CASE-II, it is clear that FL is effective in the DICE environment for ransomware spread detection. It achieved the baseline (centralized) performance both for the IID and non-IID cases. This suggests that we no longer need to compromise data privacy by using centralized learning. Moreover, we do not need sophisticated models for our cases as a simple model such as logistic regression is sufficient to detect the ransomware spread effectively.  

\subsubsection{Overhead time for the model training:}
\hskip2pt The anti-ransomware model training time in DICE is also critical as the model needs to be updated regularly to learn about the characteristics of a newly available ransomware spread. As such, we investigate the overhead time for the model development in FL and centralized baselines. 
\begin{figure} [t!]
	\centering
	\includegraphics[width=0.5\linewidth]{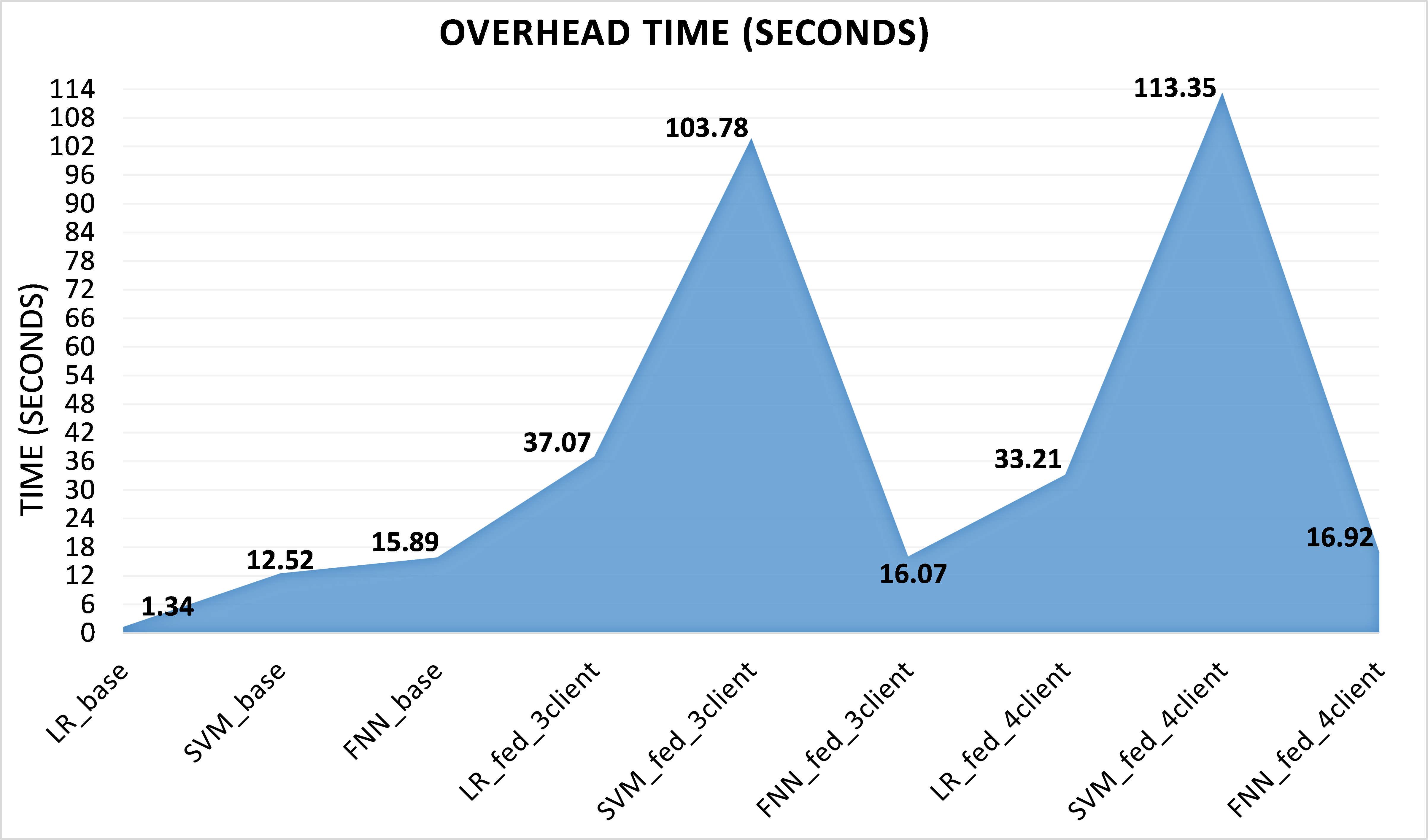}
	\caption{\small{Overhead time comparison for centralized and federated learning (CASE-I).}}
	\label{fig:training_testing_time} 
\end{figure}

Our results are depicted in Fig.~\ref{fig:training_testing_time}. It shows that FL has some higher overhead time for model processing (training and testing) than their centralized counterparts. This can be considered as the trade-off to data privacy and distributed computations. It is known that the overhead depends on the number of clients and data samples in the client; smaller data and higher clients can reduce the overhead time in FL.  

We also performed our experiments on the data with sliding window size equals 20s (besides 10s) for \textbf{RQ1} and \textbf{RQ3}. We found similar results and conclusions as above. Unfortunately, the results are excluded due to the page limitation.


\section{Experiments on the mitigation of ransomware spread} 
\label{sec:mitigation}
In this section, we present our policy-based ransomware spread mitigation technique in FedDICE. This is as an answer to \textbf{RQ4}.
%
To this end, we have created a small SDN network simulating FedDICE to test our proposed security architecture described in section~\ref{sec:proposed_solution}. For simulation simplicity, we represent the hierarchy of controllers using SDN data-plane devices, where every single data-plane device hosts an FL module. 
%
Our network setup has used four mininet simulated hosts (clients) connected to an ONOS SDN controller. In addition, we have tapped one Raspberry pi virtual machine (VM) as one of the hosts in this network. We consider this as an IoT device connected to a hospital network.
%

\begin{figure}[t]
	\centering
		\subfigure[]{
    		\includegraphics[width=0.95\linewidth, height=4.8cm]{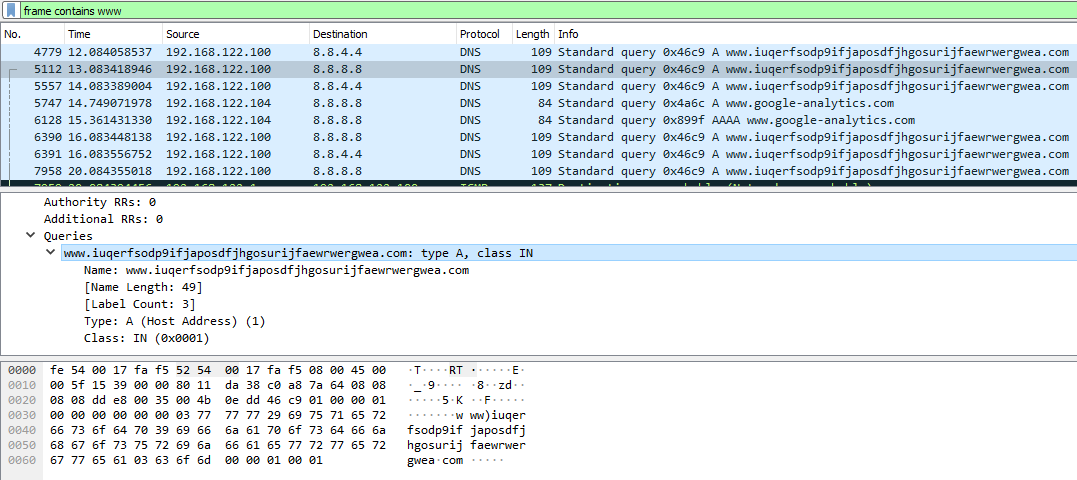}
        	\label{fig:wsCapture} 
		
		}
		
		\subfigure[]{
		    \includegraphics[width=0.28\linewidth]{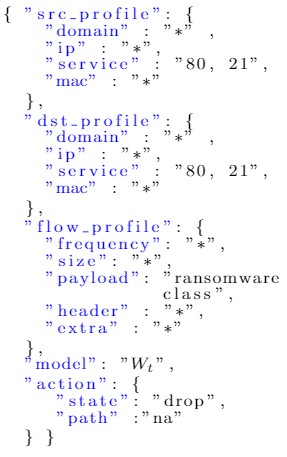}
        	\label{fig:sample_policy} 
                }
       	
		\caption{\small{(a) Wireshark capture of the communication, including that from ransomware, and (b) a sample policy in a JSON format.}}
		\label{fig:combined_results_mitigation}
\end{figure}

\textbf{Successful Attack Phase:} To simulate the Wannacry ransomware attack in our network, we have used tcpreplay~\cite{tcpreplay}. Normally, a user-driven input initiates ransomware propagation. However, for this simulation purposes, we use a previously recorded set of WannaCry packet traces. We have connected a Raspbian VM to the network and used tcpreplay to replay the packet trace. Fig.~\ref{fig:wsCapture} shows a Wireshark capture of the WannaCry communication in the network.

\textbf{Detection and mitigation phase:}  
The FL models are used as policies to detect the attacks coming towards any hospital node. Our policy syntax is granular and can extend beyond the machine learning model. The policies are extremely customizable. We are using JSON to write the policies. The formulated policies can be used in the SDN controller of any hierarchy. Together with PbSA agent running in the local ICE, the fed agent can automatically forge the policies and install them in the local controller. Also, the FL models get updated in the regular interval from the central learning node located in the supernode, allowing the other lead SDN controller to be aware of the learning model changes. Besides, each policy has multiple sections. Firstly, the source profile (src\_profile) refers to the specifics of the packets originating from the source, i.e., from which domain, IP, MAC, and service type the packet is coming from. Secondly, the destination profile (dst\_profile) talks about the specifics of the packet destination, i.e., which domain, IP, and MAC address the packet is going to. Thirdly, the flow profile (flow\_profile) specifies the flow. A series of packets constitute a flow, and each flow has a specific frequency of packet occurrence that is of specific sizes. Flows can specify payload type in the policy as well. For instance, a policy specifies the flow that needs to be dropped, e.g., flow containing ransomware packets. Fourthly, the machine learning model $W_t$ specifies the model used as the policy for checking the incoming packet flow. Finally, each policy comes with an action. An action can be as simple as dropping a flow, or it can be a complicated one that routes the flows via edge devices. For instance, we can route any sensitive healthcare device flows via trusted switching devices. We can signify a wildcard using ``*" in any policy field. Listing in Fig.~\ref{fig:sample_policy} shows a sample policy used to detect the custom ransomware attack simulated by us in the SDN infrastructure. The policy says that any packet originating from any profile targets towards any destination, which is detected as the ransomware payload by the trained model, $W_t$, then the packets will be dropped. Furthermore, the local SDN controller removes the data source from the network.

\section{Related Work}
\label{sec:related_work}
This section reviews the relevant works to cybersecurity in (distributed) ICE, ransomware detection and mitigation in clinical scenarios, and related works to federated learning in ransomware detection.  

\subsection{Cybersecurity in ICE}

An ICE is usually implemented by using OpenICE~\cite{open_ice}. Its security is managed partially by the external Data Distribution Service (DDS) middleware. However, these DDS middleware suffered from various difficulties, including complexity and scalability, in implementations~\cite{middleware}. In another work, authors have proposed a cloud-based secure logger as an effective solution against replay, eavesdropping, and injection~\cite{logger}, but the solution is ineffective if there is no message alteration.    

\subsection{Ransomware Detection}
\label{sec:ransomware_detection}
In literature, SDN-based ransomware mitigation solutions had been proposed. One of the works considered SDN redirection capabilities together with a blacklist of proxy servers to check if the infected device is trying to connect to one of them to obtain the public encryption key~\cite{cabaj2016}.
The mitigation consisted of establishing a flow filter to impede this communication, and thus, the encryption of the files. Here the main drawback is the need to keep a blacklist of proxy servers updated. These servers must be identified utilizing behavioral analysis of known malware, thus making it impossible to detect new campaigns. When compared with our solution, our goal is not to prevent the encryption of the files. We attempt to detect ransomware spreading by using the characteristic traffic patterns generated during that stage. Additionally, our mitigation procedure restores the ICE/DICE system to a clean state. In another work, a deep packet inspection was implemented to identify the HTTP POST messages from ransomware~\cite{cabaj2018}. The authors considered the lengths of each three consecutive HTTP POST messages as a feature vector and train by computing both the centroid of the feature vectors belonging to each ransomware class and a maximal distance to be considered from that class. This approach does not work if the packets are encrypted. 


\subsection{Related works to federated learning in ransomware detection}
\label{sec:fed_relatedworks}

There are two categories of related works: malware detection and anomaly detection. Various works had been done on the malware detection side, but little attention was given to ransomware detection and mitigation. Usually, these works considered host-based datasets, which are host-based systems' logs, e.g., event logs in a computer. For industrial IoT, FL was used to detect android malware applications in Industrial IoT (host-based data). Moreover, the authors considered data triggered by poisoning attacks based on GAN for robust training~\cite{fediiot}. In separate work, host-based activities log datasets of malware (including ransomware) are considered to carry FL on malware classification~\cite{fed_mal}. On the anomaly detection side, FL was implemented to detect anomalies in IoT~\cite{diot}. The data streaming from each IoT was considered and analyzed where the laptop was considered a gateway~\cite{diot}. The IoT was infected with only one malware called Mirai (not a ransomware-type). In contrast, our work has considered the network of networks beyond the gateway (e.g., laptop) and ransomware samples in a DICE environment. In separate work, FL was implemented for network anomaly detection but performed to detect VPN traffic and Tor traffic~\cite{network_anomaly}. Federated deep learning over the network flow dataset, called CIC-IDS2017, using blockchains was implemented for anomaly detection in intrusion detection. Moreover, CIC-IDS2017 did not contain data due to ransomware samples. In our work, we are focused on network security rather than an individual device. Thus, we use a dataset capturing network information (netflow). In contrast to the host-based dataset, our dataset captures completely different characteristics and belongs to a different distribution. It is unclear how the FL approach performs in a network-based ransomware dataset, so we addressed this question in this work.
\section{Conclusion and future works}\label{sec:conclusion_future_works}

In this work, we presented FedDICE, which is federated learning entangled distributed integrated clinical environment. It enables data privacy by leveraging federating learning. Our results demonstrated that FedDICE effectively detects ransomware spread detection of WannaCry, Petya, BadRabbit, and PowerGhost with a testing accuracy of around 99\% in the distributed integrated clinical environment (DICE). This performance is similar to the performance achieved by centralized learning, which is not a privacy-friendly approach. Besides, we implemented a policy-based ransomware spread mitigation technique leveraging the software-defined network functionalities. FedDICE is a generic framework and can be potentially used for other applications such as AI-supported medical decisions with data privacy. 

This work demonstrated the proof of concept of FedDICE and its application in ransomware spread detection and mitigation. Studies considering a large-scale DICE with a large number of ransomware types are interesting avenues for further explorations. Studying other threat scenarios where attackers influencing the model training, including adversarial federated learning (e.g., FedGAN~\cite{fediiot}, model poisoning attacks~\cite{survey_fed}), in the FedDICE system is left as future works. 



%
%

\bibliographystyle{splncs04}
\bibliography{mybibliography}

\begin{thebibliography}{10}
\providecommand{\url}[1]{\texttt{#1}}
\providecommand{\urlprefix}{URL }
\providecommand{\doi}[1]{https://doi.org/#1}

\bibitem{data_set}
The hospital room of the future datasets, {A}vailable:
  \url{http://perception.inf.um.es/ICE-datasets/}, last accessed on 05 Feb.
  2021

\bibitem{nist}
Nist cybersecurity framework, {A}vailable:
  \url{https://www.nist.gov/cyberframework/risk-management-framework}

\bibitem{pytorch}
Pytorch, {A}vailable: \url{https://pytorch.org/}

\bibitem{threat}
Ransomware: Past, present, and future,
  \url{https://blog.talosintelligence.com/2016/04/ransomware.html#ch3-portent}

\bibitem{tcpreplay}
tcpreplay, https://linux.die.net/man/1/tcpreplay, last accessed on 2 April 2021

\bibitem{open_ice}
Arney, D., Plourde, J., Goldman, J.M.: Openice medical device interoperability
  platform overview and requirement analysis. Biomed. Tech.  \textbf{63},
  39--47 (2018)

\bibitem{ransomware_pay}
Brok, C.: Following ransomware attack indiana hospital pays \$55k to unlock
  data (2020),
  \url{https://digitalguardian.com/blog/following-ransomware-attack-indiana-hospital-pays-55k-unlock-data#:~:text=A\%20hospital\%20in\%20Indiana\%20paid,stop\%20the\%20bleeding\%20on\%20Friday.}

\bibitem{cabaj2016}
{Cabaj}, K., {Mazurczyk}, W.: Using software-defined networking for ransomware
  mitigation: The case of cryptowall. IEEE Network  \textbf{30}(6),  14--20
  (2016)

\bibitem{cabaj2018}
Cabaj, K., Gregorczyk, M., Mazurczyk, W.: Software-define d networking-based
  crypto ransomware detection using http traffic characteristics. Computers and
  Electrical Engineering  \textbf{66},  353--368 (2018)

\bibitem{alberto_work}
Celdran, A.H., Karmakar, K.K., Marmol, F.G., Varadharajan, V.: Detecting and
  mitigating cyberattacks using software defined networks for integrated
  clinical environments. Peer-to-Peer Netw. Appl. pp. 1--16 (2021)

\bibitem{ransomware_skyrocketing2}
CheckPoint: Attacks targeting healthcare organizations spike globally as
  covid-19 cases rise again (2021),
  \url{https://blog.checkpoint.com/2021/01/05/attacks-targeting-healthcare-organizations-spike-globally-as-covid-19-cases-rise-again/}

\bibitem{miot}
Dimitrov, D.V.: Medical internet of things and big data in healthcare. Healthc
  Inform Res.  \textbf{22}(3),  156–163 (2016)

\bibitem{gdpr}
{EU}: Regulation (eu) 2016/679 general data protection regulation. Official
  Journal of the European Union  (May 2016)

\bibitem{ICE_architecture}
F2761, A.: Medical devices and medical systems - essential safety requirements
  for equipment comprising the patient-centric integrated clinical environment
  (ice) - part 1: General requirements and conceptual model. ASTM International
   (2013), \url{https://www.astm.org/Standards/F2761.htm}

\bibitem{sensor_paper}
Fernandez~Maimo, L., Huertas~Celdran, A., Perales~Gomez, Ã.L., Garcia~Clemente,
  F.J., Weimer, J., Lee, I.: Intelligent and dynamic ransomware spread
  detection and mitigation in integrated clinical environments. Sensors
  \textbf{19}(5) (2019)

\bibitem{ransomware_skyrocketing3}
Gallagher, R., Bloomberg: Hackers 'without conscience' demand ransom from
  dozens of hospitals and labs working on coronavirus (2020),
  \url{https://fortune.com/2020/04/01/hackers-ransomware-hospitals-labs-coronavirus/}

\bibitem{ML_survey}
Gibert, D., Mateu, C., Planes, J.: The rise of machine learning for detection
  and classification of malware: Research developments, trends and challenges.
  Jr. of Network and Computer Applications  \textbf{153},  102526 (2020)

\bibitem{ids_survey}
Khraisat, A., Gonda, I., Vamplew, P., Kamruzzaman, J.: Survey of intrusion
  detection systems:techniques, datasets and challenges. Cybersecur
  \textbf{2}(20),  1--22 (2019)

\bibitem{fedkonecny2}
Konecn{\'{y}}, J., McMahan, B., Ramage, D.: Federated optimization: Distributed
  optimization beyond the datacenter. arxiv  (2015),
  https://arxiv.org/pdf/1511.03575.pdf

\bibitem{middleware}
Ömer Köksal, Tekinerdogan, B.: Obstacles in data distribution service
  middleware: A systematic review. Future Gener. Comput. Syst.  \textbf{68},
  191–210 (2017)

\bibitem{fed_mal}
Lin, K.Y., Huang, W.R.: Using federated learning on malware classification. In:
  Proc. ICACT. pp. 585--589 (2020)

\bibitem{ransomware_skyrocketing}
Mathews, L.: Ransomware attacks on the healthcare sector are skyrocketing
  (2021),
  \url{https://www.forbes.com/sites/leemathews/2021/01/08/ransomware-attacks-on-the-healthcare-sector-are-skyrocketing/?sh=2c5aa87d2d25}

\bibitem{fed1}
McMahan, B., Moore, E., Ramage, D., Hampson, S., y~Arcas, B.A.:
  Communication-efficient learning of deep networks from decentralized data.
  In: Proc. AISTATS. pp. 1273--1282 (2017)

\bibitem{survey_fed}
Mothukuri, V., Parizi, R.M., Pouriyeh, S., Huang, Y., Dehghantanha, A.,
  Srivastava, G.: A survey on security and privacy of federated learning.
  Future Generation Computer Systems  \textbf{115},  619--640 (2021)

\bibitem{logger}
Nguyen, H., Acharya, B., et~al., R.I.: Cloud-based secure logger for medical
  devices. In: Proc. IEEE CHASE. p. 89–94 (2016)

\bibitem{diot}
Nguyen, T.D., Marchal, S., Miettinen, M., Fereidooni, H., Asokan, N., Sadeghi,
  A.R.: Diot: A federated self-learning anomaly detection system for iot. In:
  Proc. ICDCS. pp. 756-- 767 (2019)

\bibitem{attack1}
O'Neill, P.H.: A patient has died after ransomware hackers hit a german
  hospital (2020),
  \url{https://www.technologyreview.com/2020/09/18/1008582/a-patient-has-died-after-ransomware-hackers-hit-a-german-hospital/}

\bibitem{obfuscation}
{Riboni}, D., {Villani}, A., {Vitali}, D., {Bettini}, C., {Mancini}, L.V.:
  Obfuscation of sensitive data in network flows. In: 2012 Proc. IEEE INFOCOM.
  pp. 2372--2380 (2012)

\bibitem{naturefed}
Sheller, M.J., Edwards, B., Reina, G.A.e.a.: Federated learning in medicine:
  facilitating multi-institutional collaborations without sharing patient data.
  Scientific Reports  \textbf{10, 12598} (2020).
  \doi{10.1038/s41598-020-69250-1}

\bibitem{sherpaai}
Sherpa.ai: Federated learning framework, {A}vailable:
  \url{https://github.com/sherpaai/Sherpa.ai-Federated-Learning-Framework}

\bibitem{cyberphysical}
Stankovic, J.A.: Research directions for cyber physical systems in wireless and
  mobile healthcare. ACM Transactions on Cyber-Physical Systems  \textbf{1}(1),
   1--12 (2016)

\bibitem{fediiot}
Taheri, R., Shojafar, M., Alazab, M., Tafazolli, R.: Fed-iiot: A robust
  federated malware detection architecture in industrial iot. IEEE TII  (2020)

\bibitem{reviewpaper_our}
Thapa, C., Camtepe, S.: Precision health data: Requirements, challenges and
  existing techniques for data security and privacy. Computers in biology and
  medicine  \textbf{129},  1--23 (2021). \doi{10.1016/j.compbiomed.2020.104130}

\bibitem{verizon_report}
Verizon: Dbir 2020 data breach investigation report (2020),
  \url{https://enterprise.verizon.com/resources/reports/2020-data-breach-investigations-report.pdf}

\bibitem{data_quality}
Vogelsang, A., Borg, M.: Requirements engineering for machine learning:
  Perspectives from data scientists. In: Proc. IEEE 27th International
  Requirements Engineering Conference Workshops (REW) (2019)

\bibitem{obfuscation2}
Wang, L., Dyer, K.P., Akella, A., Ristenpart, T., Shrimpton, T.E.: Seeing
  through network-protocol obfuscation. In: Proc. of the 22nd ACM SIGSAC
  Conference on Computer and Communications Security (CCS'15). pp. 57--69
  (2015)

\bibitem{network_anomaly}
Zhao, Y., Chen, J., Wu, D., Teng, J., Yu, S.: Multi-task network anomaly
  detection using federated learning. In: Proc. SoICT. pp. 273--279 (2019)

\end{thebibliography}

%


\appendix

\section{Metrics for performance measurement}
\label{appendix:metrics}
Before introducing accuracy, precision, recall, F1-score, and false-negative ratio (FNR), we define the following terms:
\begin{itemize}
    \item True Positive (TP): Number of positive class (ransomware class) that is correctly classified as positive (ransomware class).
     \item False Positive (FP): Number of negative class (normal class) that is incorrectly classified as positive (ransomware class).
    \item True Negative (TN): Number of negative class (normal class) that is correctly classified as negative (normal class).
    \item False Negative (FN): Number of positive class (ransomware class) that is incorrectly classified as negative (normal class).
\end{itemize}
These terms are calculated from the confusion matrix as depicted in Table~\ref{tab:confusion_matrix}. In our case, labels ``0" and ``1" refer to positive (ransomware) and negative (clean), respectively. 
\begin{table}[!h]
\centering
\caption{Confusion matrix}
\begin{tabular}{ccc}
 & Positive (0) & Negative (1) \\ \cline{2-3} 
\multicolumn{1}{c|}{Ransomware (0)} & \multicolumn{1}{c|}{TP} & \multicolumn{1}{c|}{FN} \\ \cline{2-3} 
\multicolumn{1}{c|}{Clean (1)} & \multicolumn{1}{c|}{FP} & \multicolumn{1}{c|}{TN} \\ \cline{2-3} 
\end{tabular}%
\label{tab:confusion_matrix}
\end{table}

\textbf{Accuracy}: Accuracy indicates the ability of the classifier to correctly classify all classes. It is calculated as:
\begin{equation}
    \textup{Accuracy} = \frac{\textup{TP}+\textup{TN}}{\textup{TP}+\textup{TN}+\textup{FP}+\textup{FN}} \times 100 \%
\end{equation}
Its value ranges from zero to a hundred percentage, and high accuracy indicates a better classifier.

\textbf{Precision}: Precision indicates the ability of the classifier not to classify positive class if the data is actually of negative class. It is calculated as:
\begin{equation}
    \textup{Precision} = \frac{\textup{TP}}{\textup{TP}+\textup{FP}}
\end{equation}
Its value ranges from zero to one, and high precision indicates better classifier/detection from the FP perspective. 

\textbf{Recall}: Recall (also known as sensitivity) indicates the ability of the classifier to correctly classify positive class. It is calculated as:
\begin{equation}
    \textup{Recall} = \frac{\textup{TP}}{\textup{TP}+\textup{FN}}
\end{equation}
Its value ranges from zero to one, and high recall indicates better classifier/detection from the FN perspective. 

\textbf{F1-score}: F1-score is the harmonic mean of Precision and Recall. It takes both FP and FN into consideration. It is calculated as:
\begin{equation}
    \textup{F1-score} = 2\times \frac{\textup{Precision} \times \textup{Recall}}{\textup{Precision}+\textup{Recall}}
\end{equation}
Its value ranges from zero to one and high only when both precision and recall are high. It is better than accuracy as accuracy can be disproportionately skewed by the number of actual negative classes in a setup with an uneven class distribution.

\textbf{False negative rate}: False-negative rate (FNR), also known as miss rate, indicates the proportion of the positive class incorrectly classified as a negative class.
It is calculated as:
\begin{equation}
    \textup{FNR} = \frac{\textup{FN}}{\textup{FN}+\textup{TP}}
\end{equation}
Its value ranges from zero to one, and low FNR indicates better classifier/detection from the FN perspective.

\textbf{False positive rate}: False positive rate (FPR) indicates the proportion of the negative class incorrectly classified as positive class. It is calculated as:
\begin{equation}
    \textup{FPR} = \frac{\textup{FP}}{\textup{FP}+\textup{TN}}
\end{equation}
Its value ranges from zero to one, and low FPR indicates better classifier/detection from the FP perspective.
\vspace{0.5cm}
\section{Validation result for the centralized learning}
\label{appendix:val_results}

\begin{table}[h!]
\centering
\scriptsize
\caption{(\textbf{EXP-1}) Results summary for centralized learning on validating dataset (window size = 10s for the dataset).}
\begin{tabular}{@{}cccccc@{}}
\toprule
\rowcolor[HTML]{FFFFFF} 
Model & Accuracy & Precision & Recall & F1-score & FNR \\ \midrule
\rowcolor[HTML]{9AFF99} 
LR & 0.999918 & 0.999918 & 0.9999 & 0.999908 & 0.000162 \\ \midrule
SVM & 0.999365 & 0.999518 & 0.999063 & 0.99929 & 0.001873 \\ \midrule
FNN & 0.990565 & 0.99272 & 0.986325 & 0.98941 & 0.026888 \\ \bottomrule
\end{tabular}%
\label{tab:3}
\end{table}

\end{document}